\documentclass[fleqn,usenatbib]{mnras}

\usepackage{newtxtext,newtxmath}

\usepackage[T1]{fontenc}

\DeclareRobustCommand{\VAN}[3]{#2}
\let\VANthebibliography\thebibliography
\def\thebibliography{\DeclareRobustCommand{\VAN}[3]{##3}\VANthebibliography}

\usepackage{graphicx}	
\usepackage{amsmath}	
\usepackage{threeparttable}
\usepackage{verbatim}
\usepackage{xcolor}
\usepackage{geometry}
\usepackage{marginnote}
\def\msun{{\rm M_{\odot}}}

\def\be{\begin{equation}}
\def\ee{\end{equation}}
\def\le{{L_{\rm Edd}}}

\def\mbh{{M_{\rm BH}}}
\def\sgra{Sgr~A$^*$}
\def\del#1{{}}

\defcitealias{hobbs&nayakshin2009}{HN09}

\newcommand\lsim{\mathrel{\rlap{\lower4pt\hbox{\hskip1pt$\sim$}}
    \raise1pt\hbox{$<$}}}
\newcommand\gsim{\mathrel{\rlap{\lower4pt\hbox{\hskip1pt$\sim$}}
    \raise1pt\hbox{$>$}}}
    
\title[Gas cloud disruption.]{Forming Young and Hypervelocity Stars in the Galactic Centre via Tidal Disruption of a Molecular Cloud.}

\author[A. Generozov et al.]{
A. Generozov,$^{1,2}$\thanks{E-mail: al.generozov@campus.technion.ac.il}
S. Nayakshin,$^{3}$
and A. M. Madigan$^{1}$
\\
$^1$ JILA and the Astrophysical and Planetary Sciences Department, University of Colorado, Boulder, CO 80309, USA\\
$^2$ Physics Department, Technion — Israel Institute of Technology, Haifa 3200003, Israel\\
$^3$ School of Physics and Astronomy, University of Leicester, University Road, LE1 7RH Leicester, United Kingdom 
}

\date{Accepted XXX. Received YYY; in original form ZZZ}

\pubyear{2021}

\begin{document}
\label{firstpage}
\pagerange{\pageref{firstpage}--\pageref{lastpage}}
\maketitle

\begin{abstract}
The Milky Way Galaxy hosts a four million solar mass black hole, Sgr A*, that underwent a major accretion episode approximately 3-6 Myr ago. During the episode, hundreds of young massive stars formed in a disc orbiting  Sgr A* in the central half parsec. The recent discovery of a hypervelocity star S5-HVS1, ejected by Sgr A* five Myr ago with a velocity vector consistent with the disc, suggests that this event also produced binary star disruptions. The initial stellar disc has to be rather eccentric for this to occur. Such eccentric disks can form from the tidal disruptions of molecular clouds. Here we perform simulations of such disruptions, focusing on gas clouds on rather radial initial orbits. As a result, stars formed in our simulations are on very eccentric orbits ($\bar{e}\sim 0.6$) with a lopsided configuration. For some clouds counter-rotating stars are formed. As in previous work, we find that such discs undergo a secular gravitational instability that leads to a moderate number of particles obtaining eccentricities of 0.99 or greater, sufficient for stellar binary disruption. We also reproduce the mean eccentricity of the young disk in the Galactic centre, though not the observed surface density profile. We discuss missing physics and observational biases that may explain this discrepancy.
We conclude that observed S-stars, hypervelocity stars, and disc stars tightly constrain the initial cloud parameters, indicating a cloud mass between a few$\times 10^4$ and $10^5 M_{\odot}$, and a velocity between $\sim 40$ and 80 km s$^{-1}$ at 10 pc.
\end{abstract}

\begin{keywords}
hydrodynamics -- black hole physics -- Galaxy: centre
\end{keywords}



\section{Introduction}
The central parsec of our Galaxy is home to a population of young stars, consistent with a star formation episode close to the supermassive black hole (SMBH) $\sim3-6$ Myr ago \citep{lu+2013}. The young stars may be divided into two apparently distinct kinematic structures: the S-stars and the clockwise disk. The S-star cluster ($\lesssim0.04$ pc) has an isotropic and thermal eccentricity distribution \citep{gillessen+2017} and the majority appear to be B stars. Considering the strong tidal forces in S-star cluster, it is thought that its stars did not form in place, but migrated from larger scales via binary disruption \citep{hills1988,gould&quillen2003} or interaction with a gas disc \citep{levin2007}. The clockwise disc ($0.04 \sim 0.5$ pc) is so-called due to the common motion of the stars on the sky \citep{levin&beloborodov03}. The disc contains many young O and Wolf-Rayet stars in addition to B stars. Dynamical modeling yields an average stellar eccentricity of $\sim0.3$ \citep{yelda+2014}, though this measurement may be biased by binary stars as described by \citealt{naoz+2018}.
Recent observations indicate that the age of the S-stars is consistent with the age of the clockwise disc \citep{habibi+2017}, suggestive of a common origin. 
Furthermore, \citet{koposov+2019} discovered a hypervelocity star (S5-HVS1) that was ejected from the Galactic centre (GC) 4.8 Myr ago. S5-HVS1's velocity vector is also consistent with a disk origin, though there is a $\sim 10\%$ probability of a random alignment, considering the $\sim 7^{\circ}$ degree half-opening angle of the disc \citep{lu+2009}.

A parsimonious explanation for these observations is that the formation of the disc produced an episode of binary disruptions, which would leave remnant stars in the S-star cluster. The disrupted binaries may come directly from the disc, or from an older background population. The spectroscopic age of S5-HVS1 (25-93 Myr; \citealt{koposov+2019}), suggests that it was a background star excited to disruption by the disc.

\citet{madigan+2009} and \citet{generozov&madigan2020} showed that if the stars formed in a lopsided, eccentric disk, its binaries could have been excited to nearly radial orbits and tidally disrupted via the \citet{hills1988} mechanism. 
The excitation to radial orbits occurs as stellar orbits precess retrograde to their orbital angular momenta due to the influence of the surrounding star cluster. Orbits with higher eccentricities precess more slowly and end up behind the bulk of the disk, which then torques them to even higher eccentricities.
The remnant bound stars' orbits are initially highly eccentric, but approach a thermal distribution via resonant relaxation \citep{perets+2009, antonini&merritt2013,generozov&madigan2020,tep+2021}. 

Hydrodynamical simulations show that formation of a lopsided, eccentric disc can arise from the infall of a turbulent $10^4-10^5 M_{\odot}$ molecular cloud \citep{Bonnell2008, gualandris+2012} or collisions of such clouds in the vicinity of the Galactic centre (\citealt{hobbs&nayakshin2009}; \citetalias{hobbs&nayakshin2009} hereafter). Infall of smaller molecular clumps with $\sim 100 M_{\odot}$ can explain smaller compact groups of young stars in the Galactic Centre like IRS 13N \citep{jalali+2014}.

\citet{gualandris+2012} initialised their $N$-body simulations of stellar dynamics with stellar orbits from the hydrodynamical star formation simulations of \citet{Mapelli2012}. They found that eccentric disc instability indeed occurred and led to the increase of stellar eccentricities of some of the stars. However, the effect was insufficient to produce a significant number of stellar binary disruptions. \citet{gualandris+2012} noted that this result could have been due to a relatively low ($e \sim 0.2-0.4$) eccentricities of stars formed in the hydrodynamical simulations in \citet{Mapelli2012}, and that perhaps a molecular cloud on a more radial orbit could lead to a better quantitative agreement with observations.
In this paper we attempt to find cloud initial conditions that can produce more eccentric discs. We discuss the differences between our simulation setup and that in \citet{Mapelli2012} and \citet{gualandris+2012} in \S~\ref{sec:litComp}. We perform hydrodynamical simulations in which a single molecular cloud infalls onto the GC, shocks, and settles into an eccentric gaseous disc. The disc then fragments into stars. We use the orbits of the stars to initialise high precision $N$-body simulations to track close encounters with the central supermassive black hole and to make observational comparisons.

The remainder of this paper is organized as follows. In \S~\ref{sec:joint} we describe our overall approach to simulating gas cloud disruptions, using independent hydrodynamic and N-body simulations. In \S~\ref{sec:simulations} we describe the setup and results for our hydrodynamic simulations. In \S~\ref{sec:Nbody} we summarize the setup and results of our N-body simulations, and present detailed observational comparisons. In  \S~\ref{sec:discuss}, we discuss limitations of our simulations, and additional predictions. Finally, we summarize our findings in \S~\ref{sec:sum}.

\section{Joint gas dynamics and stellar N-body simulations}
\label{sec:joint}

The goal of this paper is to try and reproduce what we know about young stars in the Galactic centre from first principle simulations. Ideally, we would like to model gas dynamics, star formation, and long term evolution of stellar orbits with one code. However, this is currently impractical due to numerical limitations. Fortunately, the problem can be broken into two largely independent sub-problems: star formation and the subsequent orbital evolution. For both of these well tested numerical approaches are available.

In the first part of our numerical investigation, we perform Smoothed Particle Hydrodynamics (SPH) simulations of gas cloud infall onto the central parsec and the resulting star formation. These simulations are tailored to the relatively short gas rich epoch of $\sim (0.1-0.3)$ Myr in duration. Due to numerical expense in resolving close particle interactions, including via shocks, the SPH simulations cannot reach sufficiently close to \sgra\ to simulate highly eccentric orbits that would produce S-stars and hypervelocity stars (HVS) via the \citealt{hills1988} binary disruption mechanism. However, these simulations are sufficiently robust to resolve star formation on scales $R \gtrsim 0.01$~pc. We recall that star formation inward of this region is not in fact expected \citep{nayakshin&cuadra2005} or observed \citep{paumard+2006} because gas radiative cooling times are too long inside radius of $\sim 1"$ (0.04 pc) to permit disc fragmentation. Our gas dynamical simulations are therefore designed to capture the formation of the well known `disc stars' \citep[e.g.,][]{levin&beloborodov03} from first principles, that is together with the gas disc formation.

When at least half of the initial gas mass is turned into stars or accreted onto \sgra\ we stop these SPH simulations and use the data for the stars formed in the simulations to initialise the N-body (stars-only) simulations. The latter follow a much smaller number of particles (stars), $N\lesssim 10^3$ but are able to address orbits that happen to pass much closer to \sgra. These simulations are also run for much longer duration, $\sim 5$~Myr. Our N-body simulations are therefore streamlined to simulate -- potential -- transformation of the disc stars into S-stars via the development of a secular instability.

\section{Hydrodynamical simulations}
\label{sec:simulations}

\subsection{Background potential}\label{sec:potential}

\sgra\ is surrounded by a massive nuclear star cluster \citep{launhardt+2002}. When a massive gas cloud strikes the central parsecs of the Galaxy and gets bound to it, both \sgra\ and the cluster must respond dynamically. Readjustment of the stellar cluster and the SMBH-cluster interactions during that is a complex problem that is outside of the scope of our investigation due to numerical limitations. Observations indicate that presently \sgra\ is exactly in the centre of the Galaxy with a very small 3D velocity \citep{reid&brunthaler2020}. Therefore, for simplicity we keep \sgra\ -- a SMBH with the initial mass of $\mbh = 4\times 10^6\msun$ -- fixed in the centre of the Galaxy for the duration of our SPH simulations. The black hole is surrounded by a spherically symmetric static stellar cusp with a broken-power law profile, motivated by \citet{schodel+2018}'s fits to the surface brightness profile in the Galactic centre. In particular, the stellar density is
\begin{align}
    \rho(r) =
    \begin{cases}
    \rho_o \left(\frac{r}{3}\right)^{-1.16} & r \leq 3 {\rm pc}\\
    \rho_o \left(\frac{r}{3}\right)^{-2.9} & r > 3 {\rm pc},
    \end{cases}
    \label{eq:schodelFit}
\end{align}
where $r$ is the distance from the Galactic centre. The density is normalized so that the mass enclosed within 3 pc is $9.1\times 10^6$ $M_{\odot}$ (the mass within 1 pc is $\sim1.2\times 10^6 M_{\odot}$). We model the effects of the stellar cusp by adding an extra acceleration to each gas or star particle, viz.
\begin{align}
a= -\frac{G M_{\rm *} (r)}{r^2},
\end{align}
where $M_{\rm *}(r)$ is the stellar mass within the star's distance from the origin, $r$. We use this potential for both SPH and N-body simulations.

\subsection{Simulation physics}

Our  star formation simulations are performed with the well known cosmological hybrid $N$-body/SPH code Gadget-2 \citep{springel2005} that includes gas hydrodynamics, cooling, gravity, $N$-body dynamics, and conversion of dense self gravitating regions of gas into stars. As in \citetalias{hobbs&nayakshin2009} we use sink particle prescription to model both \sgra\ and newly born stars. In particular, the SMBH interacts with gas and stars via gravity but also accretes SPH and star particles if they approach it closer than the capture radius $r_c = 5\times 10^{-3}$~pc and if they are gravitationally bound to it. Hence the SMBH mass increases during the simulations. Stars are treated in a similar vein except their accretion radius is much smaller, $r_c = 10^{-4}$~pc, and they are of course not fixed in space, free to follow their orbits. When stars accrete SPH particles, they accrete not only mass but also momentum of the latter. 

In \citetalias{hobbs&nayakshin2009}, gas radiative cooling was modelled with an idealised `$\beta$-cooling' approach standard to simulations of self-gravitating discs \citep{rice+2005,nayakshin+2007}. This approach fixes the gas cooling time to be proportional to the local orbital period, which is useful in differentiating between fragmenting and non-fragmenting self-gravitating discs \citet{gammie2001} but is less suitable to spherical geometries. Here we use more self-consistent approach following \cite{lombardi+2015} in which radiative cooling rate is a function of the local gas density and temperature. We use the gas/dust opacity from \citet{zhu+2015}. 

When gas cools and contracts due to its own gravity sufficiently, dense self-gravitating gas condensations form. These have densities up to multiple orders of magnitude higher than the surrounding gas, and they continue to collapse, numerically speaking to a point. Star particles are introduced when gas density exceeds $\rho_{\rm sf} = 10^{-8}$~g~cm$^{-3}$. We note that $\rho_{\rm sf}$ is much higher than the local tidal density for relevant distances from SMBH, ensuring that star particles are introduced only when local gravitational collapse is well underway. Unlike \citetalias{hobbs&nayakshin2009}, we do not allow mergers between star particles to avoid potentially significant modification of their orbital elements if two star particles on very different orbits merge.

Our unit mass is $M_u = 1\msun$, unit length $L_u = 1$~pc, and unit time is $t_u = 1.49\times 10^7$ yr.

\subsection{Initial Conditions}
\label{sec:ic}

We do not have direct observational constraints on what the initial gas configuration looked like. When choosing initial conditions here, we are motivated by (1) the need to bind a significant quantity of gas to the SMBH in the centre of the Galaxy \citep[this we would not have the young stars we observe in the Galaxy's centre][]{paumard+2006}; (2) the hint provided by the eccentric disc instability mechanism for producing HVS and S-stars. For the instability to work, the initial stellar orbits need to be rather eccentric \citep{madigan+2009,gualandris+2012,generozov&madigan2020}. This obviously requires a non-axisymmetric deposition of gas into the central parsec, and strong shocks in which a large fraction of gas kinetic energy can be lost to radiation. For a single cloud falling onto \sgra this favors a cloud on a rather radial orbit, so that different sides of the cloud pass the SMBH on opposite sides, collide behind it, and then get bound to the central parsec.

Beyond these general principles there are only very rough constraints. The cloud probably originated within the inner $\sim 200$ pc since the OB disc stars appear coeval within $\sim 1-2$ Myr \citep{paumard+2006,yelda+2014}; the free-fall time of gas from distances larger than that would likely be spread over a longer time interval and hence result in a star formation event distributed over a longer time period. Exploring such initially distant gas clouds is however prohibitively expensive for us here because we wish to resolve regions as close as $\sim 0.01$~pc to the SMBH. 

We  started our investigation with a gas cloud positioned 30 pc away from the GC. This resulted in stellar discs much larger than the $\sim 0.5$~pc observed disc of young stars \citep{paumard+2006}. More reasonable results were found for a gas cloud initially a distance $R_0$=10 pc away from the Galactic centre; this is the scenario we pursue here. For simplicity we assume a uniform density spherical cloud with radius $R_c=8$ pc and mass $M_c = 10^4$, $3\times 10^4$ or $10^5 \msun$ with a velocity vector that would take the centre of the cloud to pericentre distance $R_p=0.21$ pc, if the gas self-interaction (shocks and gravity) were neglected. We found that none of the lowest mass gas clouds, $M_c = 10^4 M_{\odot}$, resulted in a sufficient mass of young stars to explain either the observed clockwise disc stars, or the S-stars, so we do not present these experiments here. Furthermore, we explored a large range of the initial magnitude of cloud velocity, $v_0$, while keeping the pericentre passage distance the same for all of the simulations. This corresponds to a positive fixed $y$-velocity component for all of our initial gas cloud configurations (see Figure \ref{fig:Cloud1_sph_small} below), and an anticlockwise motion of the cloud centre.

We found that all of the simulations that started with a relatively large initial velocity, $v_0 \gsim 100$~km/s, produced young stars on orbits distinct from those seen in the GC. Such velocities are still somewhat lower than the local escape velocity ($\sim 170$ km s$^{-1}$), so the gas is bound to the GC. However, we find that the first passage tends to bind too small a fraction of the cloud to the SMBH, and so a second approach is needed to form a strongly bound disc. During this evolution star formation begins in shocked filaments, rather than in the disc, and results in stars on eccentric orbits with semimajor axis typically greater than 1 pc. This is in contrast to the observed young stars that are contained to the central $\sim 0.5$~pc \citep{paumard+2006}. We also found that clouds on orbits with smallest initial velocities, $v_0\lesssim 20$~km s$^{-1}$, resulted in too strong cancellation of the initial angular momentum via shocks. This led to discs too small in radial extent to explain the clockwise disc. Below we therefore focus on the middle range in $v_0$ that led to star formation most reminiscent of the observed young stars in the GC. 

\subsection{Overview of SPH simulation results}

The parameters of the SPH simulations that we chose to follow up with N-body simulations are shown in Table~\ref{tab:dis2}. Here, while describing the resulting gas dynamics and star formation, we center our discussion around two numerical experiments specifically, which we refer to as Cloud 1 and Cloud 2. These two appear to reproduce observations best and are listed in the top two rows of Table~\ref{tab:dis2}.

\begin{figure*}
    \includegraphics[width=0.99\textwidth]{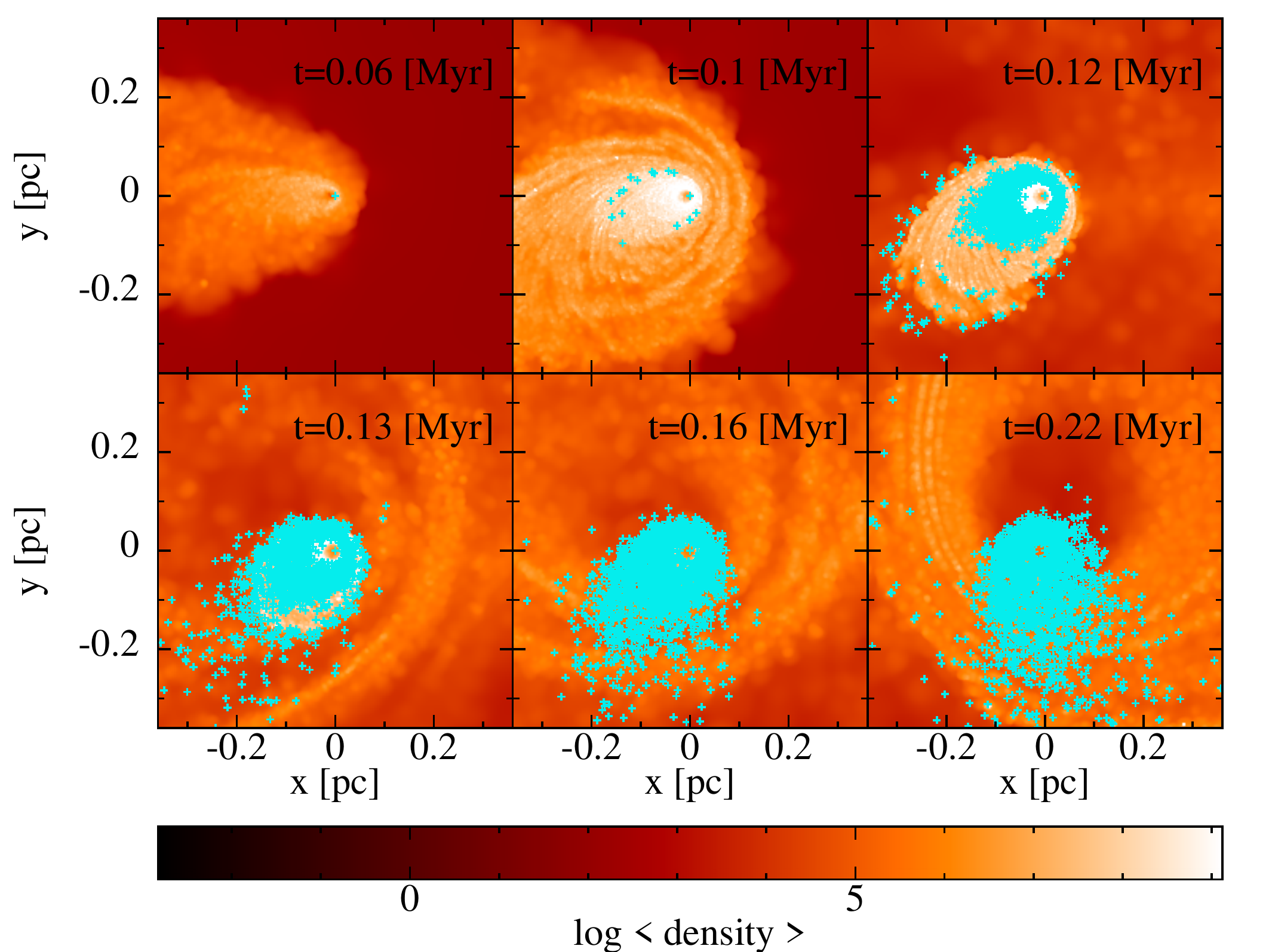}
    \caption{Snapshots of Gadget Cloud 1 simulation at selected times. Gas  density is shown with the colour map, with the colour bar in code units ($\msun$ pc$^{-3} \approx 7\times 10^{-23}$~g cm$^{-3}$). Cyan pluses mark positions of the stars formed in the simulation. The position of \sgra\ is indicated with the `+' symbol at $(x,y) = (0,0)$.  Note that gas orbits evolve from clockwise oriented at early times to anti clockwise at late times. See text for discussion of this effect.  
    }
    \label{fig:Cloud1_sph_small}
\end{figure*}

Figure~\ref{fig:Cloud1_sph_small} shows several snapshots exemplifying the complex dynamics of gas and its gradual conversion into stars in the Cloud 1 simulation. The last snapshot shown in the figure is used to initialise N-body simulations in the next section. We see that early on (in the $t= 0.06$ Myr snapshot in Figure~\ref{fig:Cloud1_sph_small}) gas with clockwise rotation dominates disc formation in the central $\sim 0.3$~pc. The disc is very eccentric (with typical eccentricity $e\sim 0.6$) and is not smooth because it forms from gas falling towards SMBH along a shocked filament located outside the figure, at $y\approx 0$ along large negative $x$. As gas infall into the centre continues, some of the filaments making up the disc become dense enough to be self-gravitating. This results in the formation of self-bound, high density clumps inside the filaments, some of which can be seen as nearly white coloured clumps in the $t=0.1$ Myr snapshot. When the star formation density threshold is exceeded in these clumps, star particles are introduced. These are shown with the cyan crosses (the very central cross is \sgra\ itself). 

Since stars are born from the gas, they inherit the orbital parameters of the parent SPH particles, and initially follow very similar orbits. However, new gas arriving in the central region constantly mixes with the gas already there. Thus, gaseous orbits evolve under the effects of both gravity and shocks and/or gas pressure forces whereas star evolve mainly due to gravity.\footnote{Dynamical friction and gas accretion onto stars are relatively weak compared to gravitational forces acting on stars in the central parsec} This leads to a gradual divergence in orbits of stars and gas. 

One can also see in the top row of snapshots shown in Figure~\ref{fig:Cloud1_sph_small} some apsidal orbital precession for both gas and stars due to the stellar potential in the GC. However, the bottom row of the snapshots shows strikingly different orbital structure for stars and gas. There is a depletion of gas in the central region, with most of the mass there being in the young stars (not counting the background potential, of course). Furthermore, as time progresses, the sign of the angular momentum of the dominant inflow of gas into the central region flips. By the time of the last snapshot shown the gas rotates predominantly counterclockwise in the snapshot. This very rapid evolution is due to star formation, strong shocks, and angular momentum cancellation, an effect that we shall discuss in greater detail below.

\begin{figure*}
    \centering
    \includegraphics[width=0.99\textwidth]{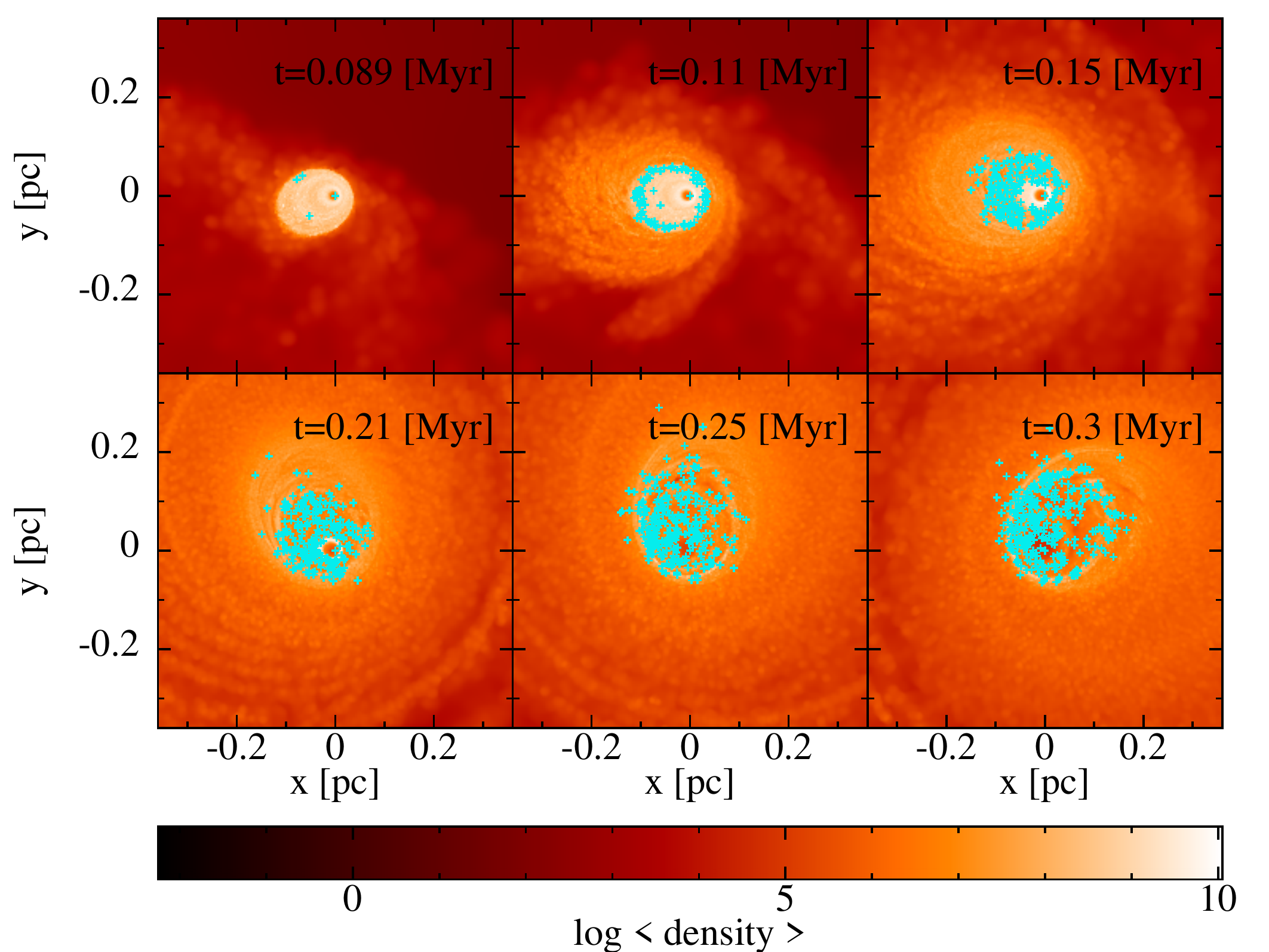}
    \caption{Same as Figure~\ref{fig:Cloud1_sph_small} but for Cloud 2. Note that the orientation of the dominant component of gas in the very centre is opposite to that of Cloud 1 at all times. Therefore the stars that form there orbit \sgra in the counterclockwise direction.}
    \label{fig:Cloud2_sph_small}
\end{figure*}

Figure~\ref{fig:Cloud2_sph_small} shows gas and star dynamics in the central region for Cloud 2 simulation. The orientation of angular momentum vector for the gas dominating the infall at all times is counterclockwise, so we do not observe a flip in the angular momentum vector as in the Cloud 1 simulation. The forming disc is still eccentric but somewhat less so, and is also more uniform. This indicates that shocks are less important in the central $\sim 0.2$~pc for the Cloud 2 simulation than they were in Cloud 1 simulation. As a result, gravity dominates the dynamics of both gas and stars in the central region, and thus gaseous and stellar orbits in the Cloud 2 simulation are much closer to one another than in Cloud 1 simulation. The last snapshot shown in Figure~\ref{fig:Cloud2_sph_small} is used to start our $N$-body simulations. 

\begin{figure*}
    \includegraphics[width=0.99\textwidth]{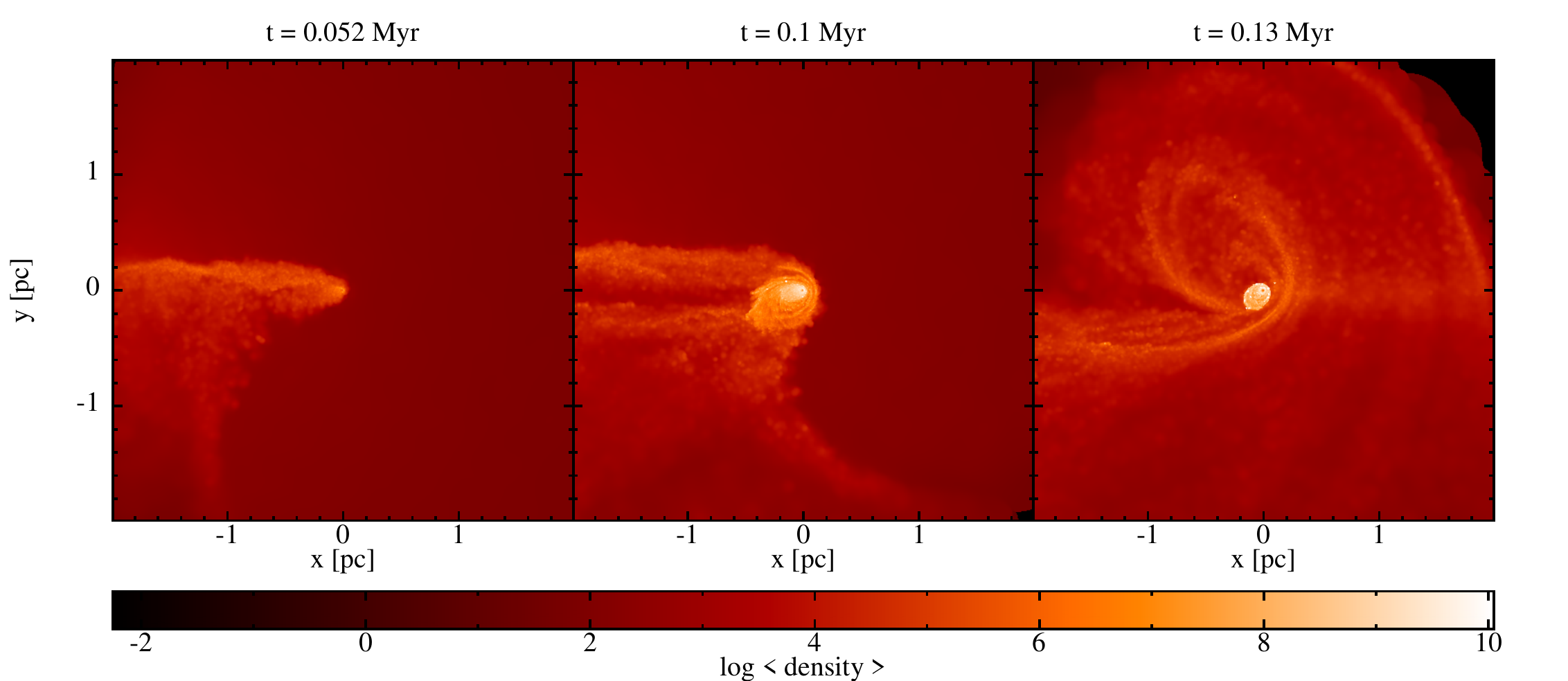}
    \includegraphics[width=0.99\textwidth]{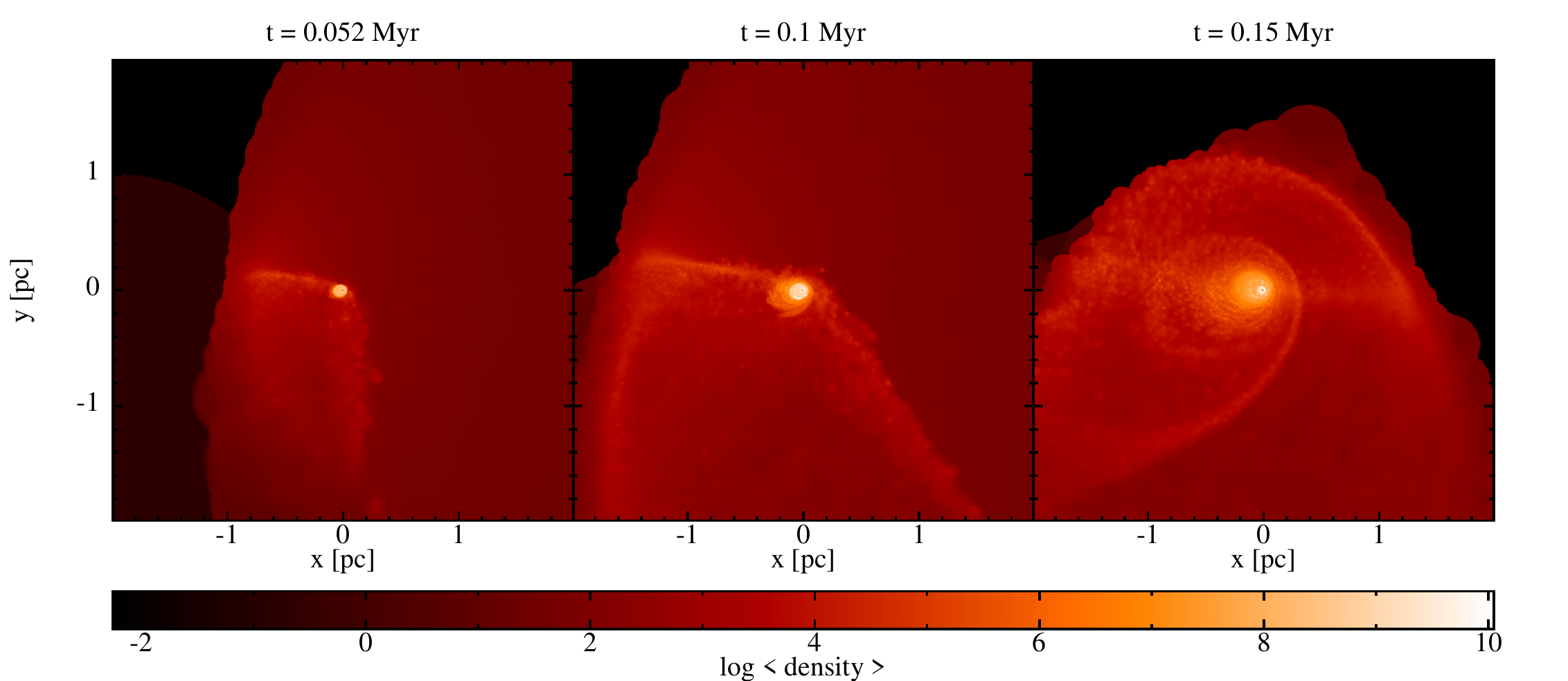}
    \caption{Cloud 1 (top) and Cloud 2 (bottom) at selected times (0.052, 0.1 and 0.13 Myr from left to right, respectively) on larger physical scales than shown in \ref{fig:Cloud1_sph_small} and \ref{fig:Cloud2_sph_small}. Note that in Cloud 1 simulation the gas inflowing into the very central region is dominated by the filament oriented along the negative $x$ axis and has a clockwise angular momentum.}
    \label{fig:Clouds_largeR}
\end{figure*}

Figure~\ref{fig:Clouds_largeR} compares gas dynamics of Cloud 1 (top row of panels) and Cloud 2 (bottom row of panels). In the former case gas entering the inner regions ($\sim 0.25$~pc) has a clockwise angular momentum at early times (the leftmost panel). By the time $t=0.1$ Myr (middle panel), there are two gas streams entering the central region, one with clockwise angular momentum and one with anticlockwise angular momentum. Finally, at later times (the right panel), there is only one dominant orientation of angular momentum for the incoming gas: it is now anticlockwise. Note that the anticlockwise direction is the dominant one in the cloud, so it `wins' eventually in Cloud 1 simulation. In the Cloud 2 simulation, however, we observe that the clockwise filament feeding the central region is never dominant. Instead it is the anticlockwise one that always dominates gas supply to the central disc. 

\begin{figure}
    \includegraphics[width=0.95\columnwidth]{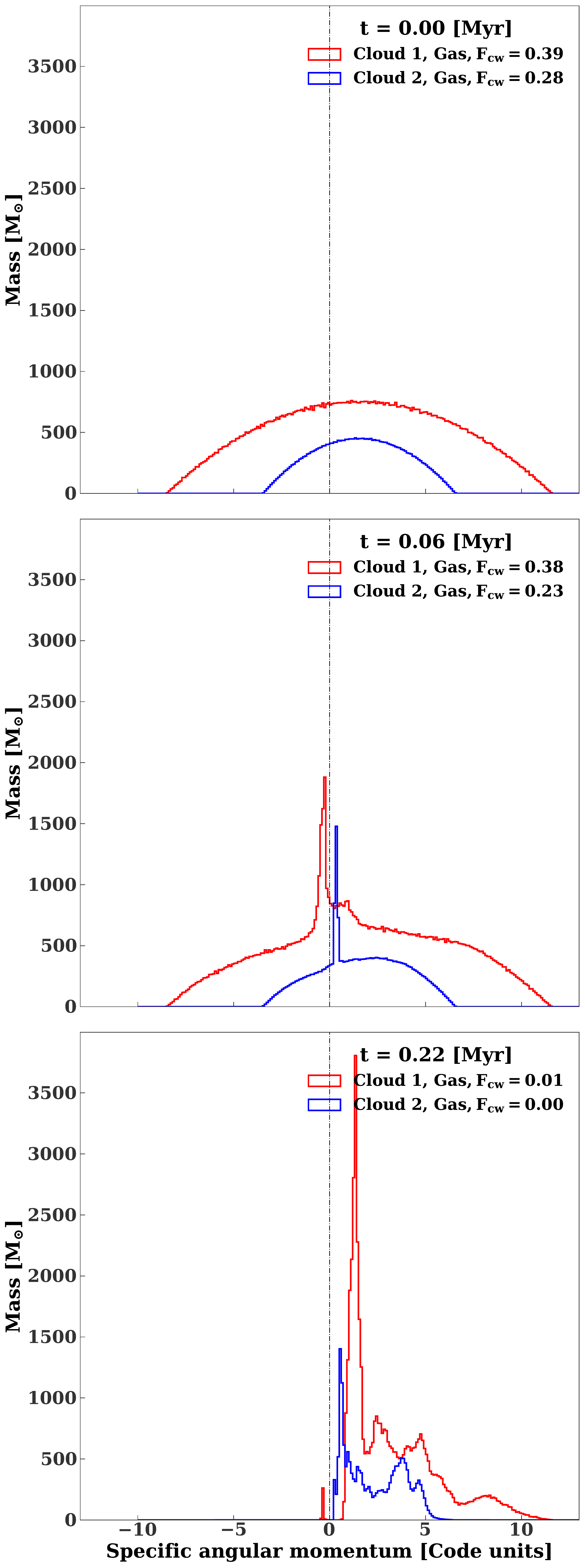} \caption{Specific angular momentum distribution of gas in Cloud 1 and Cloud 2 at a few different snapshots. For each cloud and snapshot we indicate `$F_{\rm cw}$'--the fraction of particles on clockwise orbits with negative angular momentum.}     \label{fig:Gas_Lz}
\end{figure}

\begin{figure}
    \includegraphics[width=0.95\columnwidth]{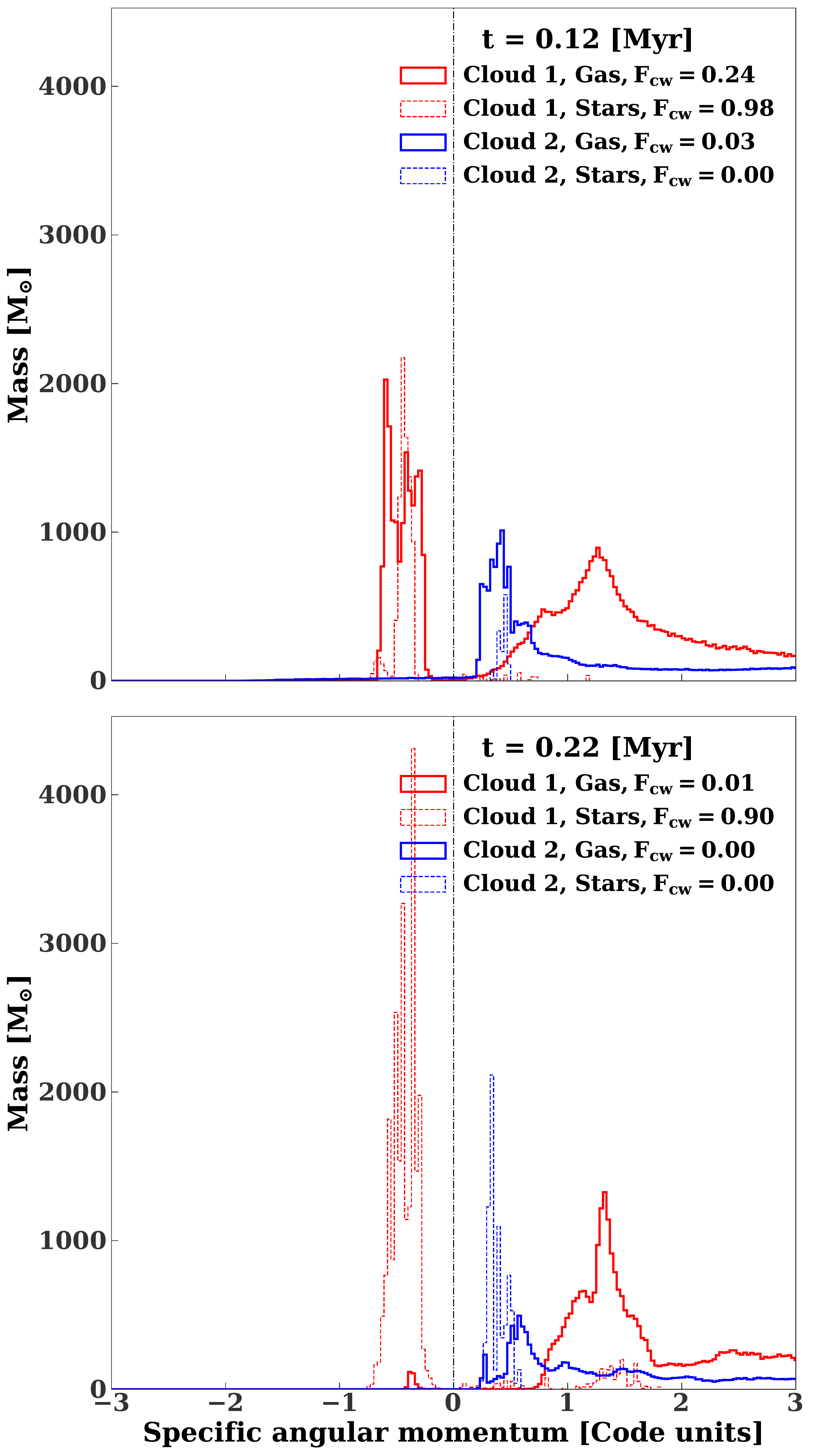}
        \caption{Specific angular momentum distribution of star particles formed from Cloud 1 and 2 (thin lines). For reference, we also plot the distribution of gas at the same times with thick lines.}
        \label{fig:Star_Lz}
\end{figure}

Figure~\ref{fig:Gas_Lz} presents angular momentum distribution for SPH particles in the simulations for Cloud 1 and Cloud 2 at several times. The dashed vertical line marks purely radial orbits ($j_z = 0$). Note that the mean of $j_z$ at $t=0$ is exactly the same for two clouds, but because the 3D initial velocity of Cloud 1 is larger, the initial distribution of angular momenta for Cloud 1 is broader than that for Cloud 2. The legend lists the fraction of SPH particles that have `clockwise' orbits (with negative $j_z$), $F_{\rm cw}$.

The higher $F_{\rm cw}$ for Cloud 1 is very important. At $t=0.06$~Myr we observe low angular momentum peaks for both Clouds 1 and 2, but they have opposing signs. These peaks are formed due to shocks. As streams with positive and negative angular momentum collide, the absolute value of angular momentum, $|j_z|$ is reduced for both streams. The exact location of the peak in $j_z$ is however not always trivially predicted since it depends on which of the streams gets into \sgra\ vicinity first, their relative densities and the amount of mass and momentum present in both. We see from Figure~\ref{fig:Gas_Lz} that in Cloud 1 the negative $j_z$ material has the upper hand but only at early times. Eventually, as more and more gas with positive angular momentum falls in, we see that the negative shoulder of $j_z$ distribution disappears in both Clouds 1 and 2.

However, this -- temporary -- victory of negative $j_z$ gas in the inner disc has a lasting effect on the resulting stellar distribution. Figure~\ref{fig:Star_Lz} shows the distribution of angular momenta for stars formed in these two simulations at several different times. We can see that the stars in Cloud 1 are mainly rotating clockwise, whereas those in Cloud 2 all rotate anticlockwise. Note that if we continued the SPH simulations for longer, more stars in Cloud 1 with anticlockwise orbits could have formed, but they are likely to be outside of the central $\sim 0.5$~pc that we focus on here given their relatively large angular momentum.

\subsection{Accretion onto \sgra}
As shown in the fifth column of Table~\ref{tab:dis2}, \sgra accretes between $\sim 5\times 10^3$ and $4\times 10^4 M_{\odot}$ of stars and gas ($16$ to $42\%$ of the mass of the original cloud) over the course of our simulations.

This accretion can explain the Fermi Bubbles--kpc-scale gamma ray structures, extending above and below the Galactic centre \citep{su+2010}. The bubbles indicate \sgra was active within the last few Myr, and output $10^{55}-10^{57}$ erg of energy during this period \citep{guo&mathews2012}. All of the clouds in Table~\ref{tab:dis2} can account for this energy, as long as the radiative efficiency exceeds $\sim0.1\%$.

\section{$N$-body simulations}
\label{sec:Nbody}

It is necessary to switch to a high accuracy $N$-body integrator to dynamically evolve the stellar disc going forward. Binaries that undergo disruption by the SMBH move along high eccentricity orbits which require short, adaptive timesteps to minimize energy and angular momentum error particularly near pericenter. We must also do away with the capture radius of the SMBH in the hydrodynamical code to retain high eccentricity orbits at pericenter. 

We  select snapshots from the hydrodynamic simulations that formed well-defined disks for follow-up $N$-body simulations. In particular, we select snapshots where roughly half of the initial gas mass has either been accreted by \sgra or turned into stars. The full set of cloud masses, velocities and snapshot times we used for follow-up $N$-body simulations are listed in Table~\ref{tab:dis2}. 

The positions and velocities of the stars and central SMBH are used to initialize \texttt{REBOUND} $N$-body simulations \citep{rein.liu2012}. We exclude unbound stars and those with semimajor axes greater than 0.5 pc. (We don't expect the outer regions of the disc to be dynamically relevant, considering the short range of secular torques; \citealt{gurkan&hopman2007}). For computational efficiency, we remove any binaries from the initial conditions by deleting one star from each binary pair. We ignore the remaining gas from the hydrodynamical simulations in our $N$-body simulations. This is in contrast to \citet{gualandris+2012} who include in their simulations an axisymmetric potential modelled on the remnant gas mass profile. The dynamics of the stars is dominated by the more massive spherically symmetric stellar cusp and we note that \citet{gualandris+2012} found similar dynamical evolution in simulations that excluded the gas disc potential.
\citet{subr&haas2016} also ran $N$-body simulations of an eccentric disc in the galactic center, motivated by \citet{Mapelli2012}, but did not include the stellar cusp which is crucial for comparing to observational data. 

To quickly identify the most promising Clouds for producing binary disruptions and the S-stars, we use low resolution simulations. Instead of using all of the stars, we initialize five \texttt{REBOUND} simulations with different random sets of 100 stars from each snapshot.\footnote{In practice, we first select a random set of 150 stars from these snapshots, delete all binaries, and then delete more stars until the total remaining is 100.} As discussed in \S~\ref{sec:dis}, we then select the most promising Clouds (the aforementioned Clouds 1 and 2) for follow-up, higher resolution simulations. We find excellent agreement for the number of disruptions produced by high and low resolution simulations, though we find the eccentricity distribution and surface density profile of the disc are resolution dependent, as discussed in \S~\ref{sec:bulkProp}.

The mass of the central SMBH is $\sim 4\times 10^6 M_{\odot}$, and is taken directly from the hydrodynamic simulations. The disc mass in the $N$-body simulation is the total mass of all bound stars with semimajor axes less than 0.5 pc in the corresponding Gadget snapshot ($2.2\times 10^4 M_{\odot}$ and $8.1\times 10^3 M_{\odot}$ for Clouds 1 and 2 respectively). Stars are initialized with the same mass in the $N$-body simulations.
Each disc is evolved forward in time for 6 Myr using the IAS15 integrator \citep{rein.spiegel2015}. Perturbations from the surrounding stellar cusp are included with the \texttt{REBOUNDX} package \citep{tamayo+2019}.

Our $N$-body simulations are initialized with the center-of-mass of the black hole-disc system at the origin. However, in the presence of the external cusp force the center-of-mass can drift. To be consistent with our hydrodynamic simulations (where the central black hole is fixed), we keep the the centre-of-mass of the black hole and disc fixed at the origin. 

We also performed a few simulations, where the black hole and disc are allowed to drift. In this case, the SMBH feels the acceleration of the stellar cusp if it moves more than $5\times 10^{-4}$ pc from the origin, where the enclosed stellar mass is $\approx 1 M_{\odot}$. The results are consistent with simulations where the center-of-mass drift is negated.

Unlike the simulations of \citet{generozov&madigan2020}, the simulations here do not include post-Newtonian corrections. This is because, simulations with post-Newtonian effects can have significant numerical artifacts. For example, the supermassive black hole can receive spurious velocity kicks and start drifting from the center. Such kicks occur when particles penetrate within tens of gravitational radii of \sgra, where the first order post-Newtonian prescription in \texttt{REBOUNDX} is inadequate. In most post-Newtonian simulations, such kicks do not occur, and we find good agreement with Newtonian simulations (see also \citealt{wernke&madigan2019}).

\subsection{Results of N-body simulations:
 Disruptions}
\label{sec:dis}
    Table~\ref{tab:dis2} shows statistics on `binary disruptions' in our simulations. 
    The initial conditions do not have any binaries, as they are deleted for computational efficiency before the start of the simulation. However, we record a disruption whenever particles cross within $3\times 10^{-4}$ pc of the central SMBH.\footnote{This is approximately the tidal radius of a 20 $M_{\odot}$ binary with a semimajor axis of 1 AU.} We do not delete such particles (`disruptees') from our simulations, though only a single disruption is recorded for each one. We also do not include disruptees in plots of orbital elements at late times. In reality, binaries would typically be split into two stars that are removed from the disc (with one deposited at small semimajor axes and one ejected from the system). To test the effect of this removal, we ran additional simulations, where disruptees were merged with the central black hole. We found this did not affect the disruption statistics.
    
    \begin{table*}
\caption{\label{tab:dis2} Cloud simulations that form well-defined stellar disks, and are chosen for follow-up N-body simulations. The first and second rows correspond to Clouds 1 and 2, respectively, the most promising initial conditions for reproducing observations of the Galactic centre.}
\begin{threeparttable}
\begin{tabular}{cllllllll}
Cloud mass & Cloud velocity & Snapshot time\tnote{1} & Disc mass & Accreted Mass\tnote{2} & <e> & $f(j/j_{\rm lc}<10)$\tnote{3} & Percent Disrupted\tnote{4} & Percent Disrupted\tnote{5}\\
$M_{\odot}$ & km s$^{-1}$ & yr & $M_{\odot}$ & $M_{\odot}$ & &  &  & (High resolution)\\
\hline \hline
(C1) $   1 \times 10^{ 5 }$ & 82 & $ 2.2 \times 10^{ 5 }$ & $ 2.2 \times 10^{ 4 }$ & $2.65\times 10^4$ & 0.55 & 0.5 & 7-8 [7.8] & 7.8\\
(C2) $   3 \times 10^{ 4 }$ & 41 & $ 3.0 \times 10^{ 5 }$ & $ 8.1 \times 10^{ 3 }$ & $4.71\times 10^3$ & 0.55 & 0.55 & 4-8 [5.6] & 5.5\\
\hline
$   1 \times 10^{ 5 }$ & 54 & $ 3.0 \times 10^{ 5 }$ & $ 5.5 \times 10^{ 3 }$ & $4.11\times 10^4$ & 0.63 & 0.045 & 0-0 [0.0] & \\
$   1 \times 10^{ 5 }$ & 54 & $ 3.7 \times 10^{ 5 }$ & $ 1.1 \times 10^{ 4 }$ & $4.13\times 10^4$ & 0.56 & 0.05 & 0-1 [0.2] & \\
$   1 \times 10^{ 5 }$ & 54 & $ 4.5 \times 10^{ 5 }$ & $ 1.3 \times 10^{ 4 }$ & $4.15\times 10^4$ & 0.53 & 0.094 & 0-0 [0.0] & \\
$   1 \times 10^{ 5 }$ & 54 & $ 5.4 \times 10^{ 5 }$ & $ 1.3 \times 10^{ 4 }$ & $4.16\times 10^4$ & 0.48 & 0.08 & 0-0 [0.0] & \\
$   3 \times 10^{ 4 }$ & 54 & $ 4.5 \times 10^{ 5 }$ & $ 2.9 \times 10^{ 3 }$ & $1.15\times 10^4$ & 0.62 & 0.036 & 0-0 [0.0] & \\
$   3 \times 10^{ 4 }$ & 54 & $ 6.0 \times 10^{ 5 }$ & $ 3.4 \times 10^{ 3 }$ & $1.17\times 10^4$ & 0.58 & 0.096 & 0-0 [0.0] & \\
$   3 \times 10^{ 4 }$ & 68 & $ 4.2 \times 10^{ 5 }$ & $ 5.1 \times 10^{ 3 }$ & $1.03\times 10^4$ & 0.58 & 0.27 & 1-2 [1.2] & \\
$   3 \times 10^{ 4 }$ & 82 & $ 3.0 \times 10^{ 5 }$ & $ 6.6 \times 10^{ 3 }$ & $8.84\times 10^3$ & 0.57 & 0.47 & 2-7 [4.2] & \\
$   3 \times 10^{ 4 }$ & 82 & $ 3.7 \times 10^{ 5 }$ & $ 6.7 \times 10^{ 3 }$ & $8.94\times 10^3$ & 0.54 & 0.48 & 1-3 [2.4] & \\
\hline
\end{tabular}
\begin{tablenotes}
\item [1] Starting time for N-body simulations.
\item [2] Mass accreted by \sgra during SPH simulations.
\item [3] Fraction of particles that start with angular momentum within a factor of 10 of the loss cone.
\item [4]  Percent of disc particles that become disruptees over 5 Myr from five different 100-particle simulations. The number in bracket is the mean.
\item[5] Percent of disc particles that become disruptees in high resolution simulations (see text for details).
\end{tablenotes}
\end{threeparttable}
\end{table*}

    We have identified two promising snapshots for producing binary disruptions, where $6-8\%$ of the particles became disruptees, corresponding to the first and second row of Table~\ref{tab:dis2} (Cloud 1 and Cloud 2). These two clouds produce enough disruptions to explain the bulk of the S-star population, as discussed below.
    
    Following \citet{generozov&madigan2020}, we expect
     \begin{equation}
        N_s=f_{\rm mf} N_d f_d (1-f_c) 
    \end{equation}
    S-stars within the observed mass range, where $N_d$ is the total number of binaries in the disk, $f_d$ is the fraction that are disrupted, $f_c$ is the fraction of binaries that collide instead of disrupting, and $f_{\rm mf}$ is the fraction of S-stars within the observed mass range ($\sim 3-15 M_{\odot}$). Assuming $f_d\approx 0.06-0.08$, as it is for Clouds 1 and 2, and assuming $N_d=1000$, $f_c=0.2$, and $f_{\rm mf}=0.33$ as in \citet{generozov&madigan2020}, we would predict 21 (16) S-stars between $3$ and $15 M_{\odot}$ for Cloud 1 (Cloud 2). The collision probability, $f_c$, may be a factor of $\sim 2-3$ higher once repeated close encounters with the SMBH are taken into account \citep{generozov&perets2021}. On the other hand, $N_d$ could also be higher by a factor of a few. 
    Observationally, there are $\sim$20 early-type S stars with semimajor axes less than 0.03 pc \citep{gillessen+2017}.

\begin{figure}
        \includegraphics[width=\columnwidth]{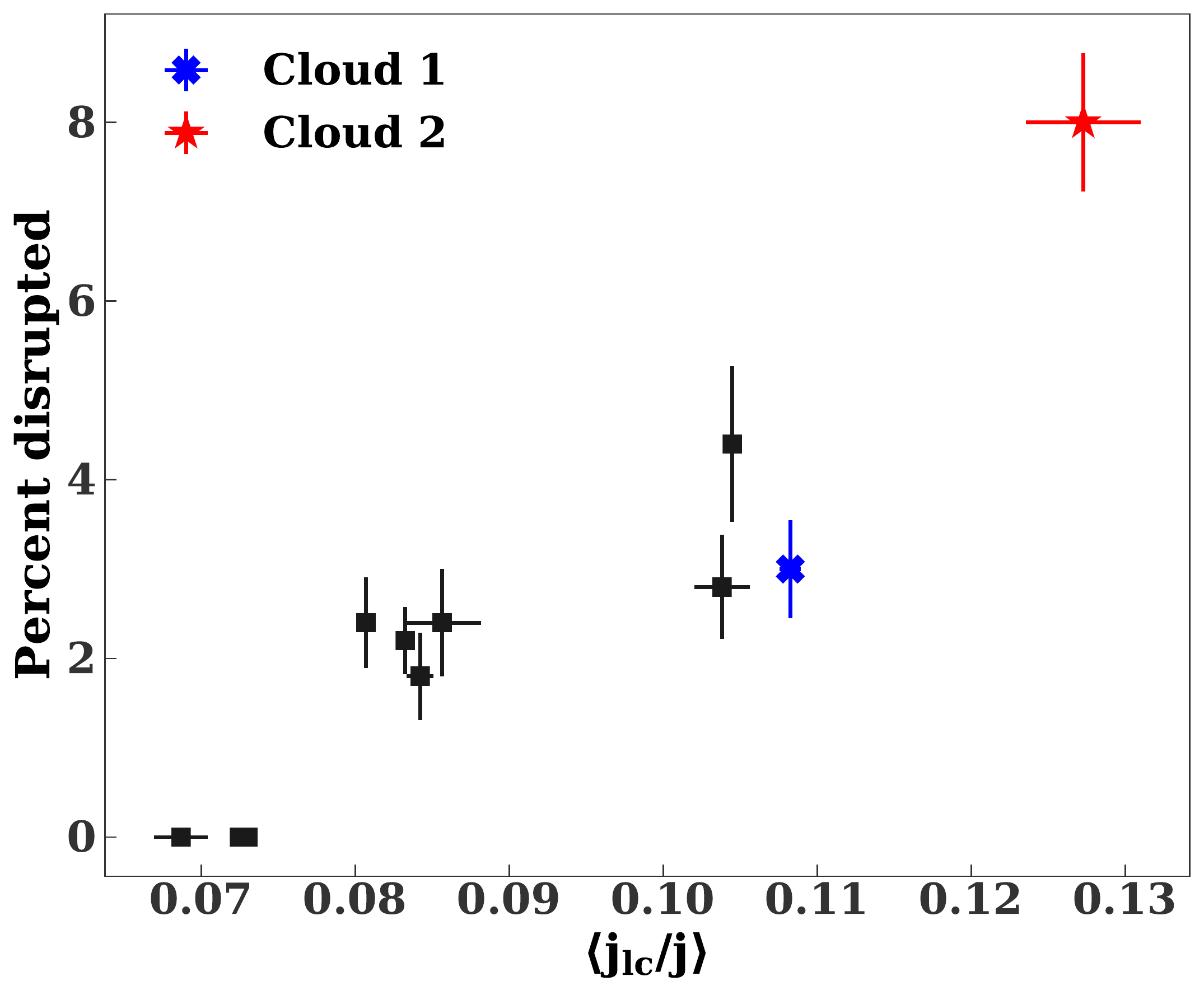}
        \includegraphics[width=\columnwidth]{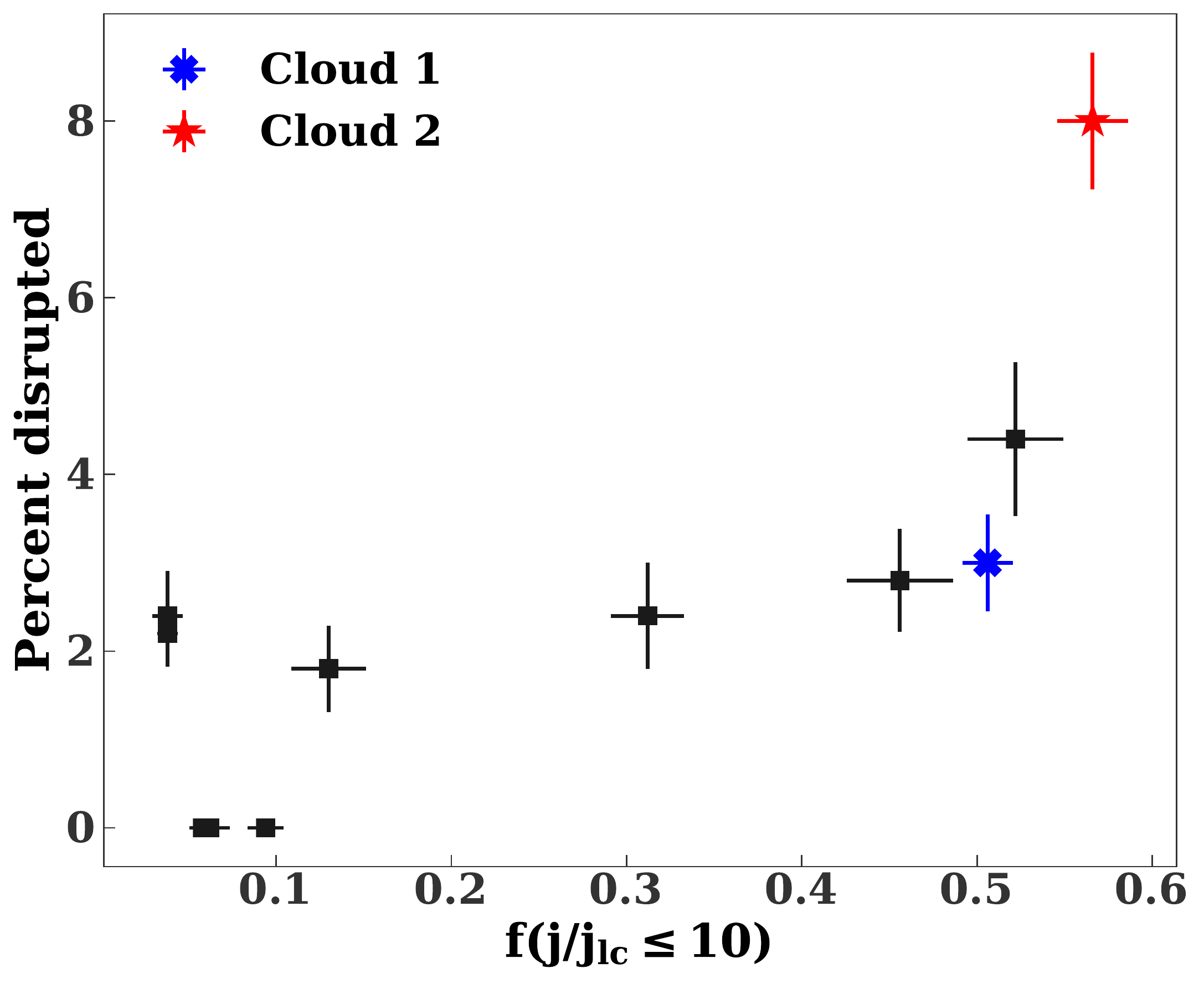}
        \caption{Mean number of disruptions for different clouds, as a function of the harmonic mean of the stars' angular momentum (\emph{top}) and the fraction of stars within a factor of 10 of the loss cone (\emph{bottom}). The clouds are the same as those in Table~\ref{tab:dis2}, except the masses are fixed to $10^4 M_{\odot}$ to better isolate the effect of the angular momentum distribution. Clouds with more stars at low angular momenta have more disruptions. This data is from low resolution (100-particle) simulations. Error bars correspond to the standard error of the mean.}
        \label{fig:jlcPlot}
    \end{figure}

     The other simulations in Table~\ref{tab:dis2} produced compact eccentric disks, but few disruptions.
     In some cases this can be explained by the small mass of the stellar disc, as some snapshots have stellar discs with only a few$\times 10^3 M_{\odot}$. We also find a correlation between the the fraction of stars at small angular momentum and the number of disruptions, as shown in Figure~\ref{fig:jlcPlot}. The top panel of this figure shows the number of disruptions versus the harmonic mean of the angular momentum (in units of the loss cone angular momentum) for different clouds. (We use the clouds in Table~\ref{tab:dis2}, but fix the stellar mass to $10^4 M_{\odot}$ in each case to better isolate the effect of the angular momentum distribution). As the fraction of stars near the loss cone increases, the harmonic mean and the number of disruptions also increase. In the bottom panel of Figure~\ref{fig:jlcPlot} we show the number of disruptions as a function of the fraction of stars within a factor of 10 of the loss cone. There is a positive correlation between the number of disruptions and this fraction too.

       So far, we have shown results from stacked 100-particle simulations. To check whether small particle numbers are affecting our results, we ran higher resolution simulations of Clouds 1 and 2 (For Cloud 1 the higher resolution simulation had 900 stars after the binary deletion step, which is approximately half the number of bound stars within 0.5 pc in the corresponding hydrodynamic snapshot. For Cloud 2 there were 328 stars, close to the total number within 0.5 pc).
       As shown in the final column of Table~\ref{tab:dis2}, the fraction of disruptees is independent of resolution. Thus, we expect our prediction for the fraction of binaries disrupted to be robust.
       
       However, as discussed in the following section, the semimajor axis and eccentricity distribution is resolution dependent, with the disc becoming more spread out and eccentric in lower resolution simulations, due to artificially strong two-body relaxation.

\subsection{Eccentricity and semimajor axis distribution} 
\label{sec:bulkProp}
\begin{figure*}
    \includegraphics[width=\columnwidth]{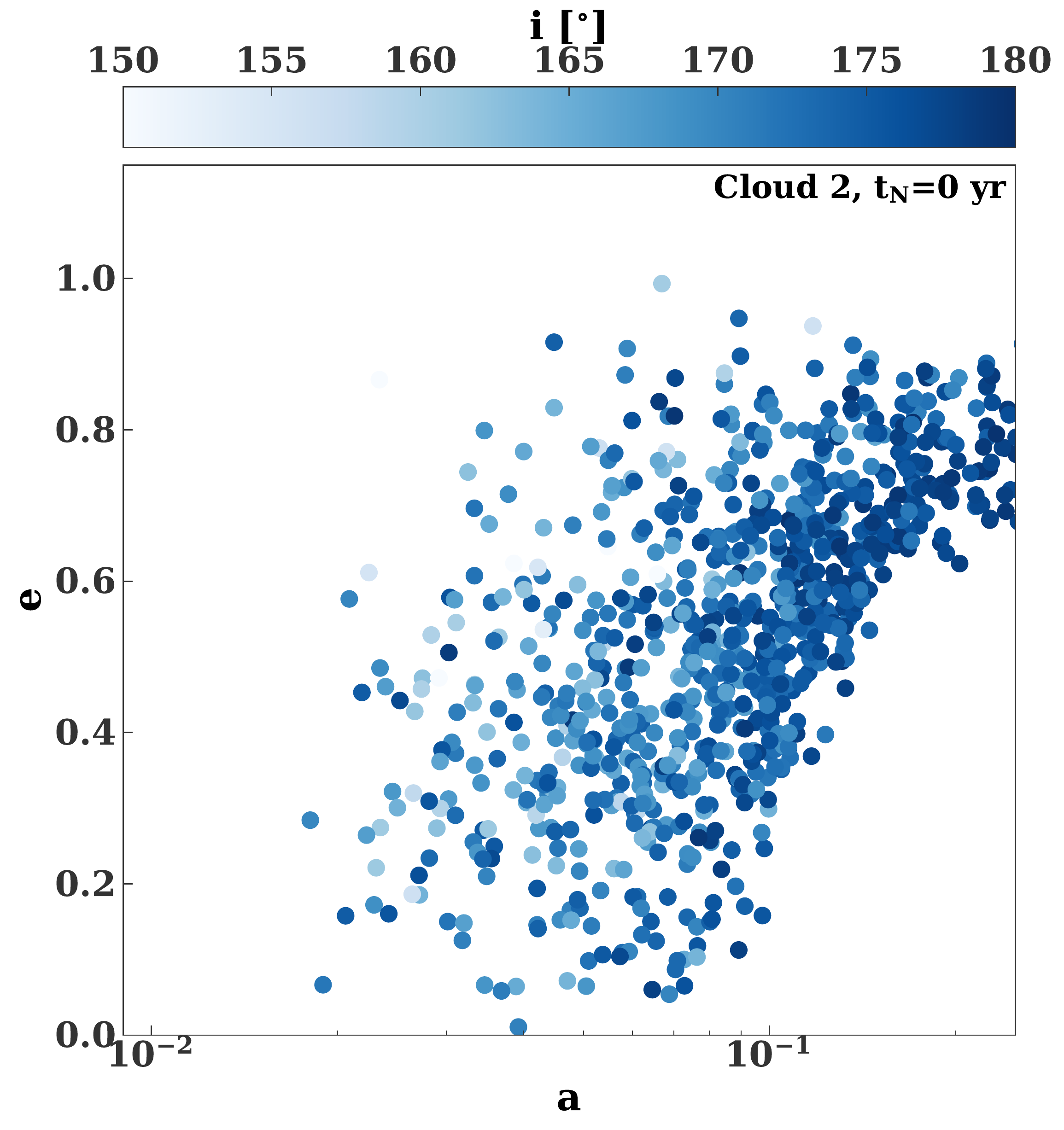}
    \includegraphics[width=\columnwidth]{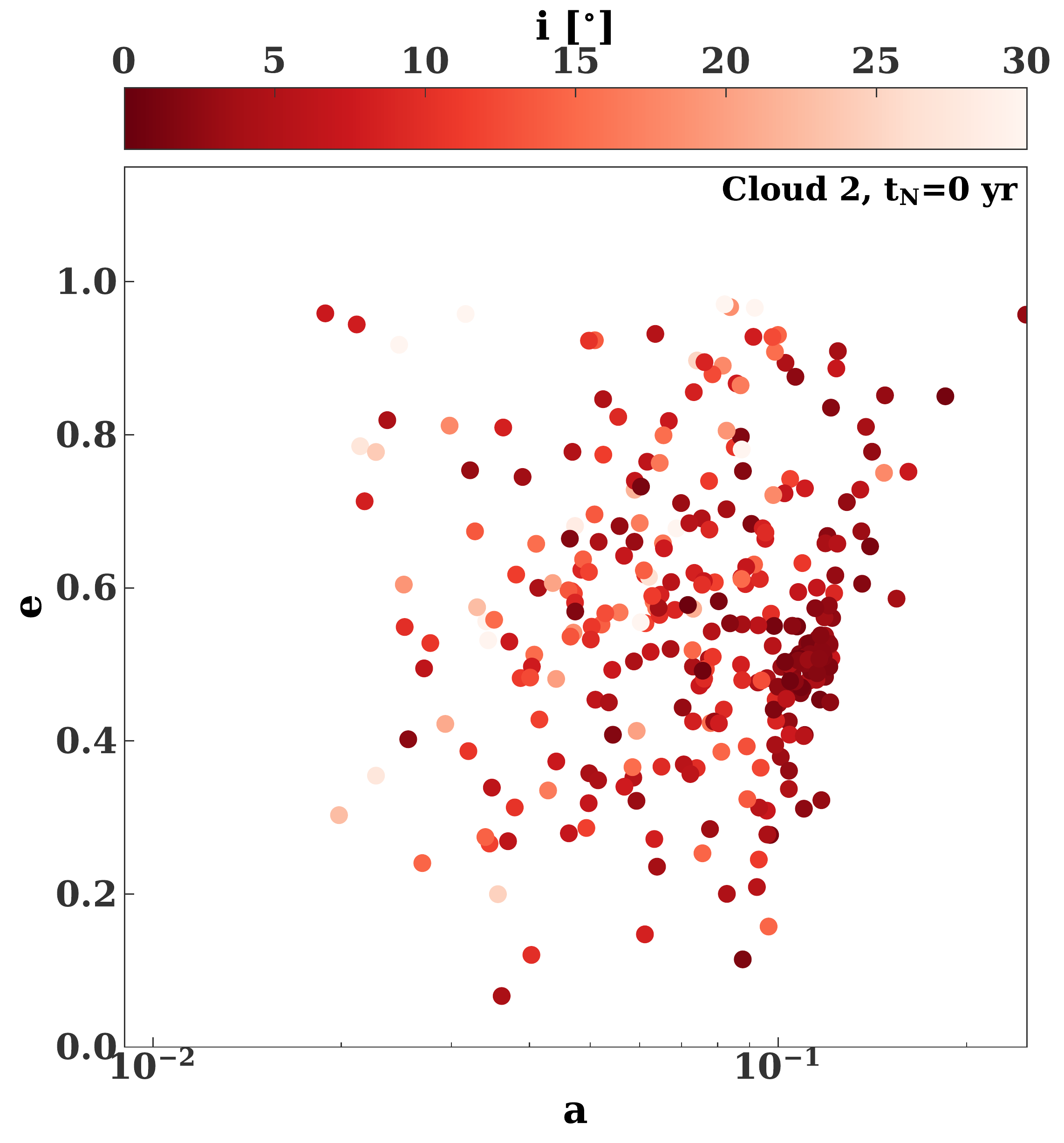}
    \caption{Initial profile of stellar orbital eccentricity versus semimajor axis from high resolution Cloud 1 (left) and Cloud 2 (right) simulations. Colors show inclinations.}
    \label{fig:before}
\end{figure*}

Figure~\ref{fig:before} shows the initial semimajor axis and eccentricity distribution for the high resolution Cloud 1 and 2 simulations. (Note that we use $t_N$ to denote time from the start of our N-body simulations). Initially, stars formed in each cloud have a similar spread of semimajor axes and eccentricities.\footnote{Here we use osculating orbital elements about the central supermassive black hole.} The outer stellar orbits in Cloud 1 are initially more eccentric than the inner orbits. This positive eccentricity gradient is caused by more efficient gas-driven cicularization in the inner disk.

\begin{figure*}
    \includegraphics[width=0.91\columnwidth]{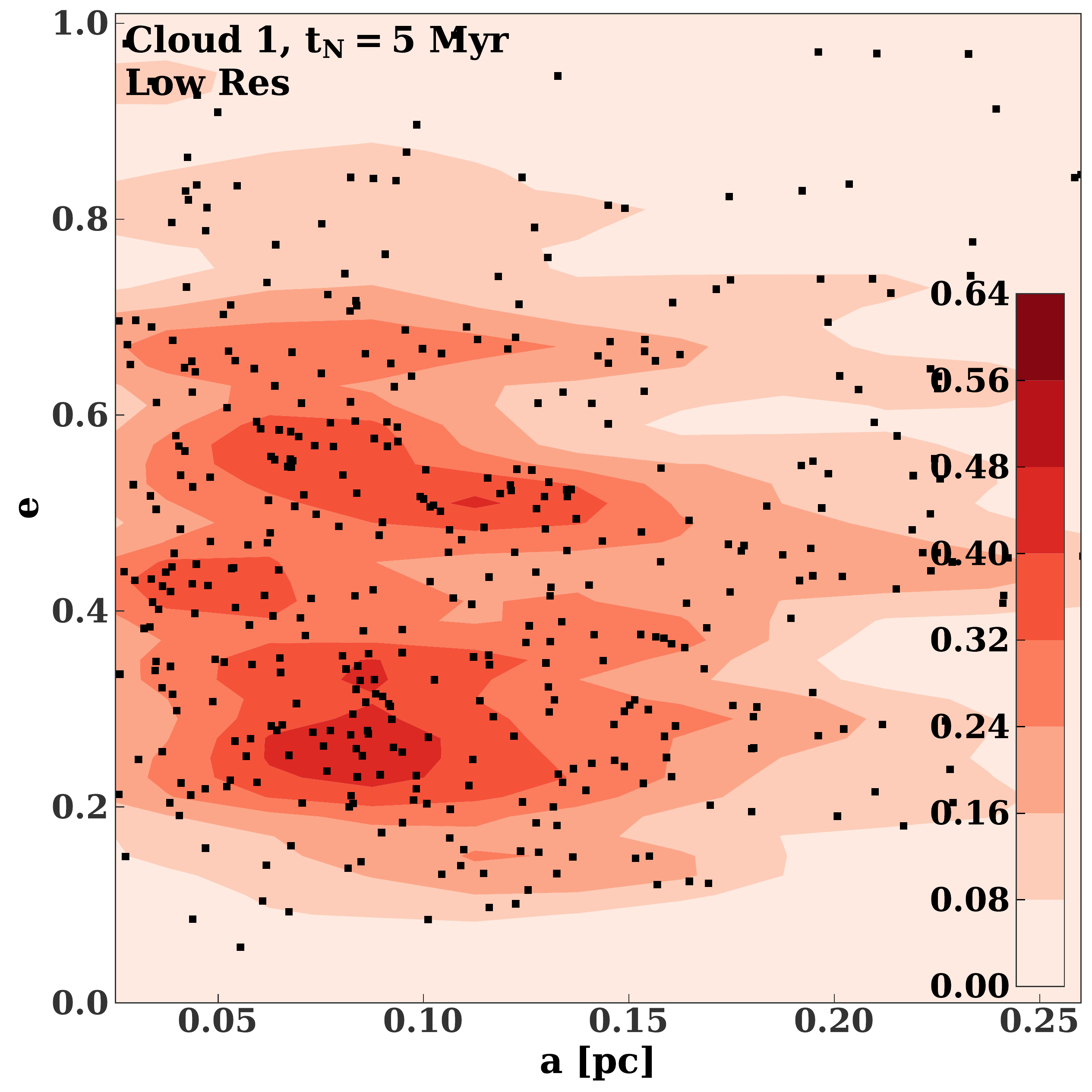}
    \includegraphics[width=1.09\columnwidth]{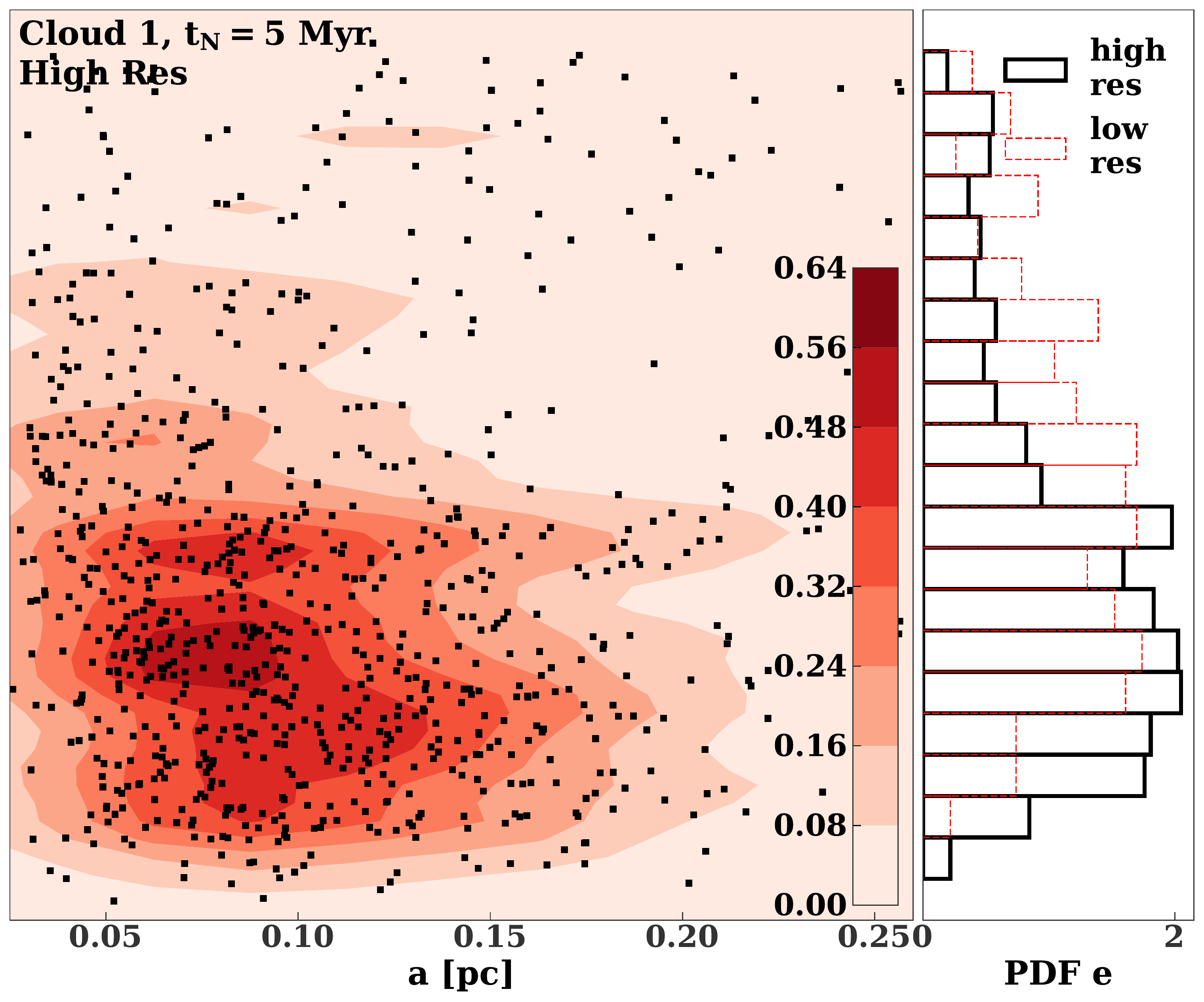}
    \includegraphics[width=0.91\columnwidth]{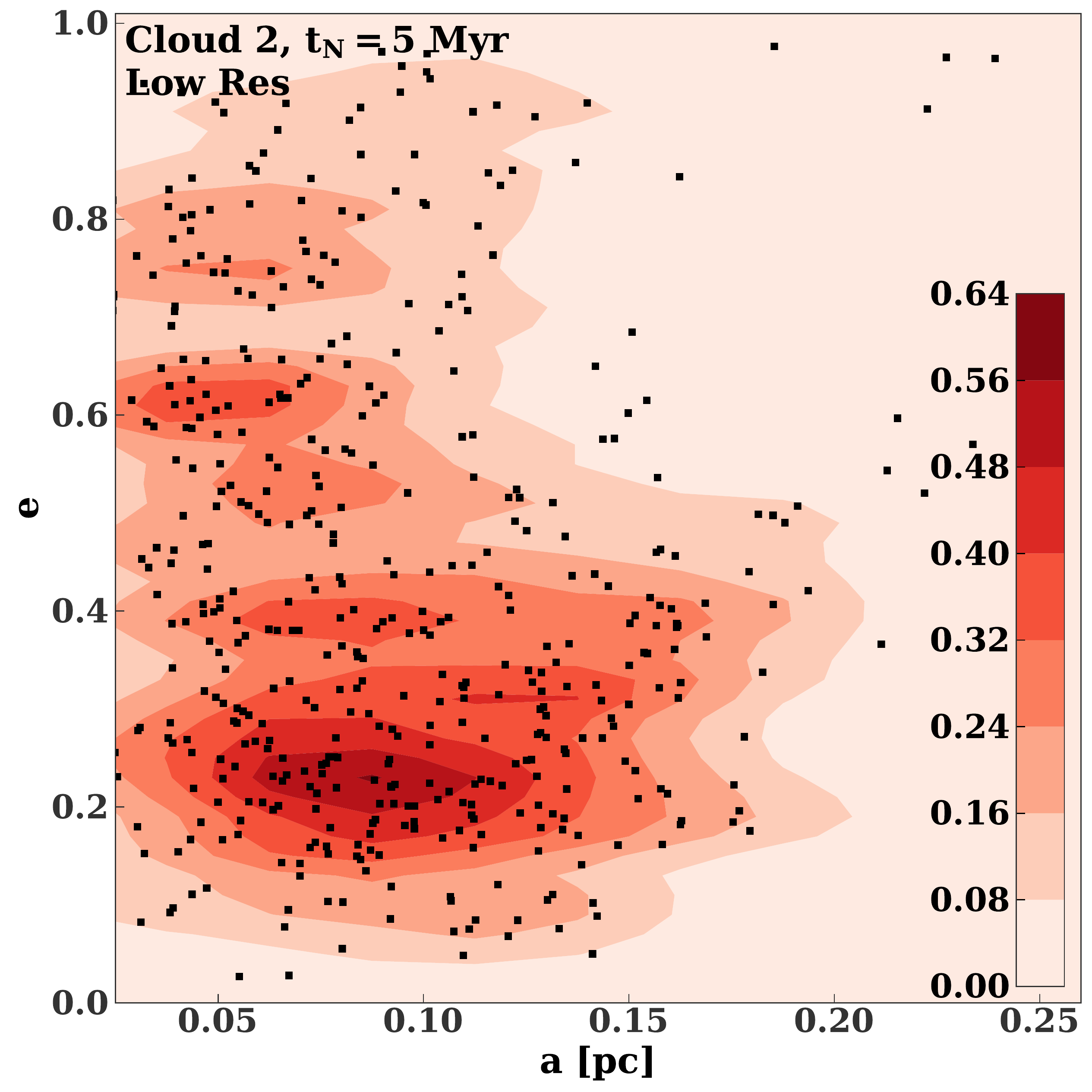}
    \includegraphics[width=1.09\columnwidth]{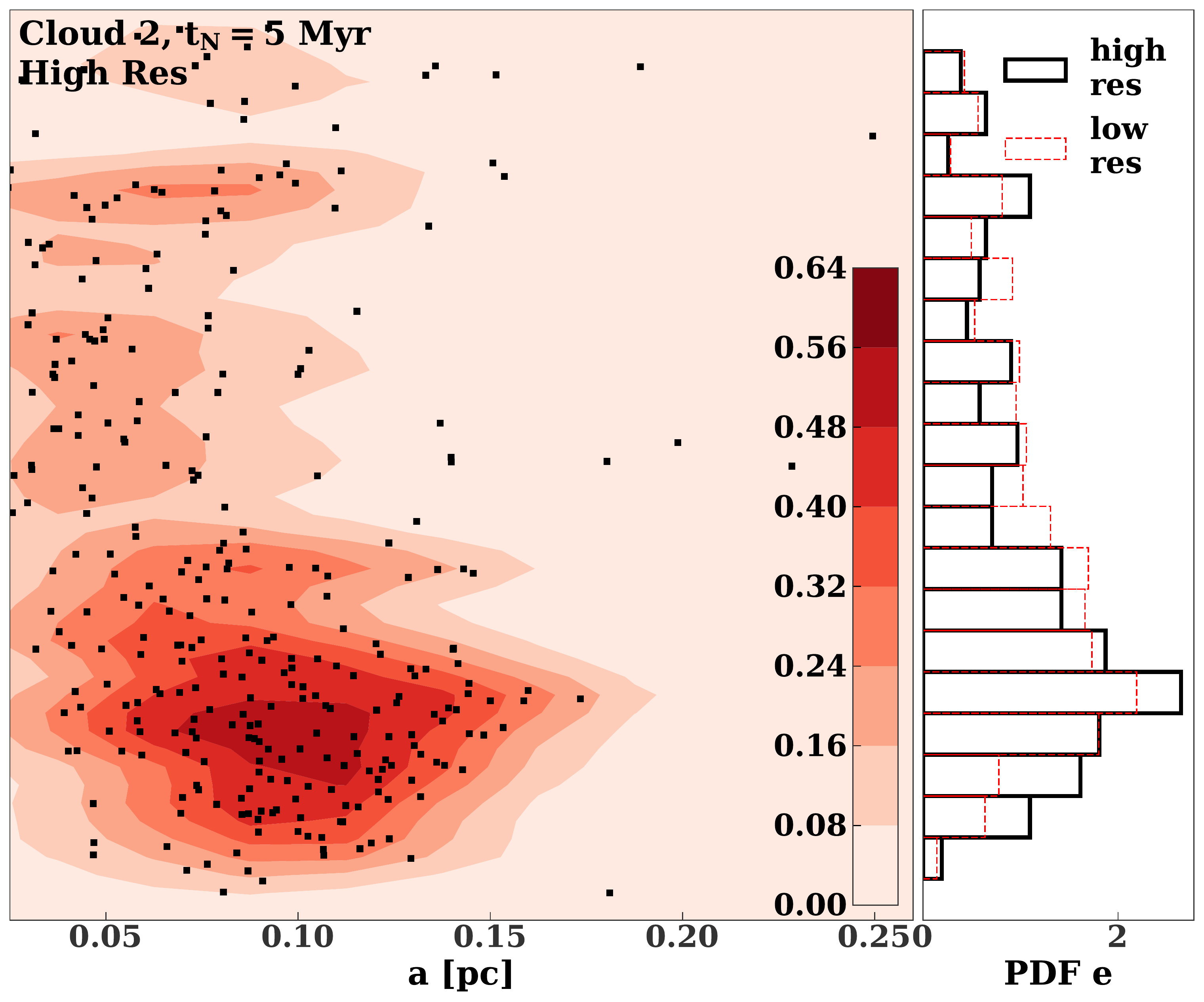}
    \caption{\label{fig:resComp} Eccentricity versus semimajor axis after 5 Myr for Cloud 1 (top row) and Cloud 2 simulations (bottom row). The left and centre panels show the distribution from low and high resolution simulations, respectively. In the former case, we use stacked data from five different simulations. The right panels compare the eccentricity distributions from these simulations. Black points are individual particles from the simulations. Red contours show the particle density in this space, and are constructed by binning the black points and smoothing with a Gaussian kernel.
    }
\end{figure*}

\begin{figure*}
    \includegraphics[width=\columnwidth]{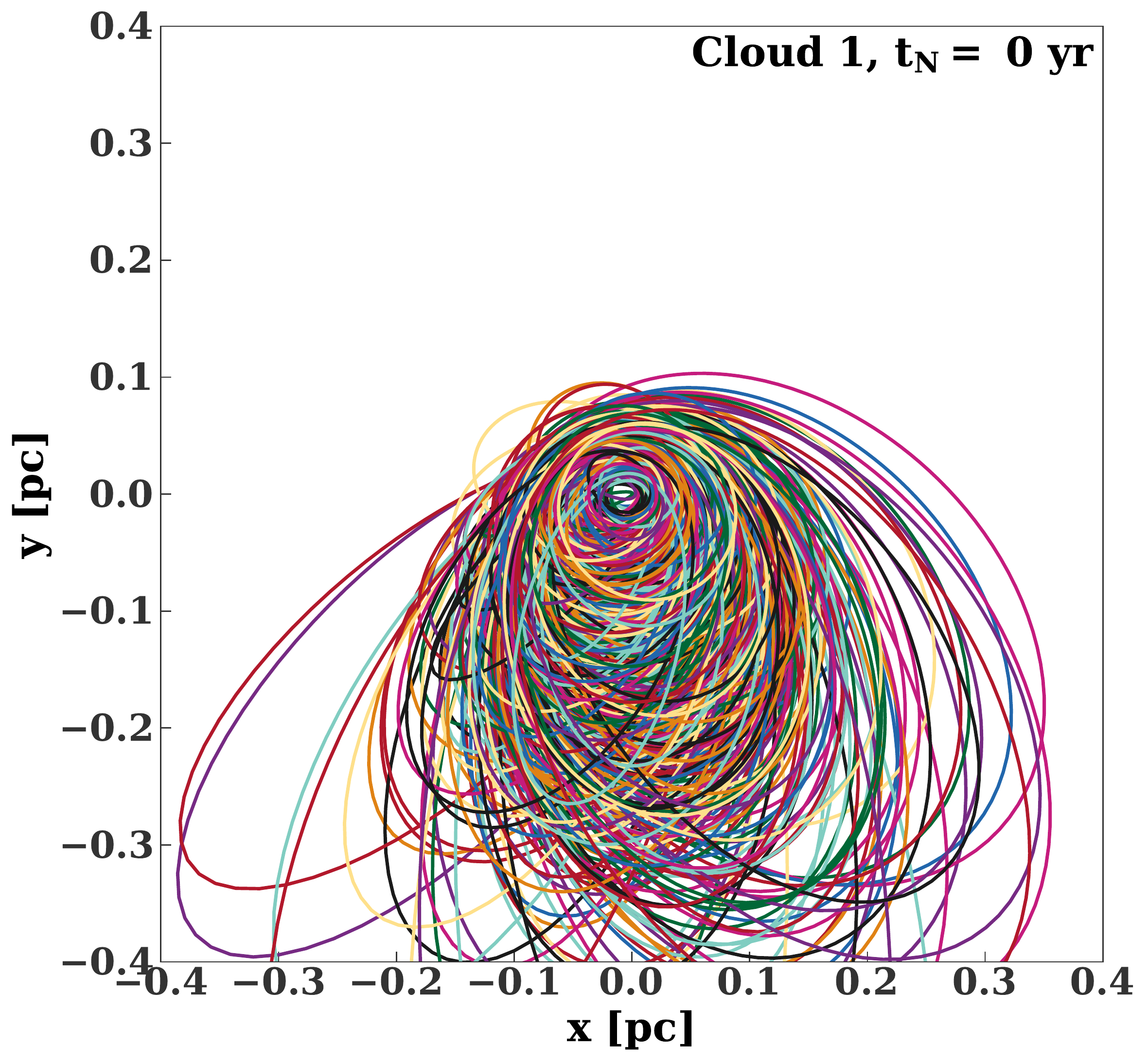}
    \includegraphics[width=\columnwidth]{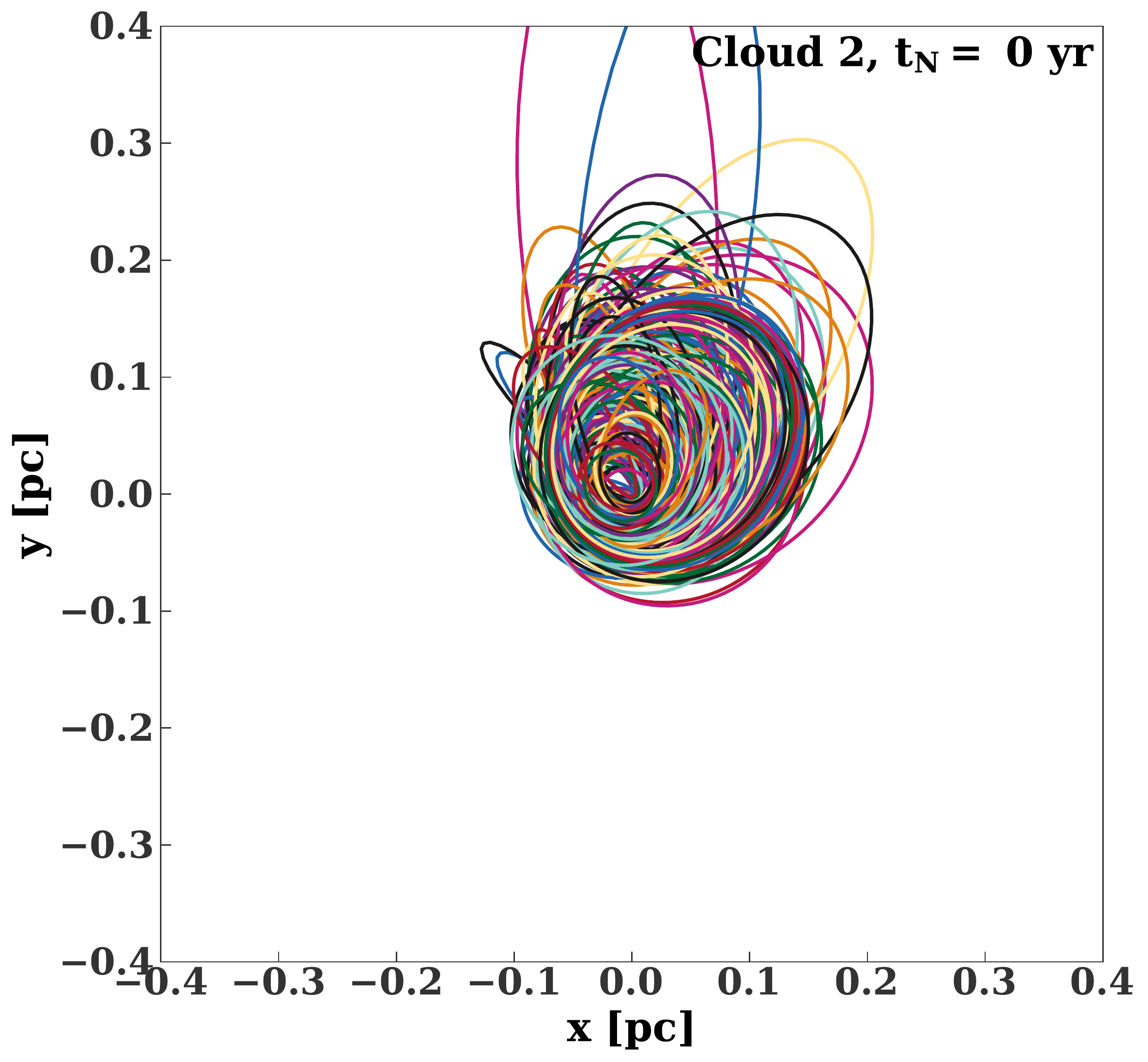}
    \includegraphics[width=\columnwidth]{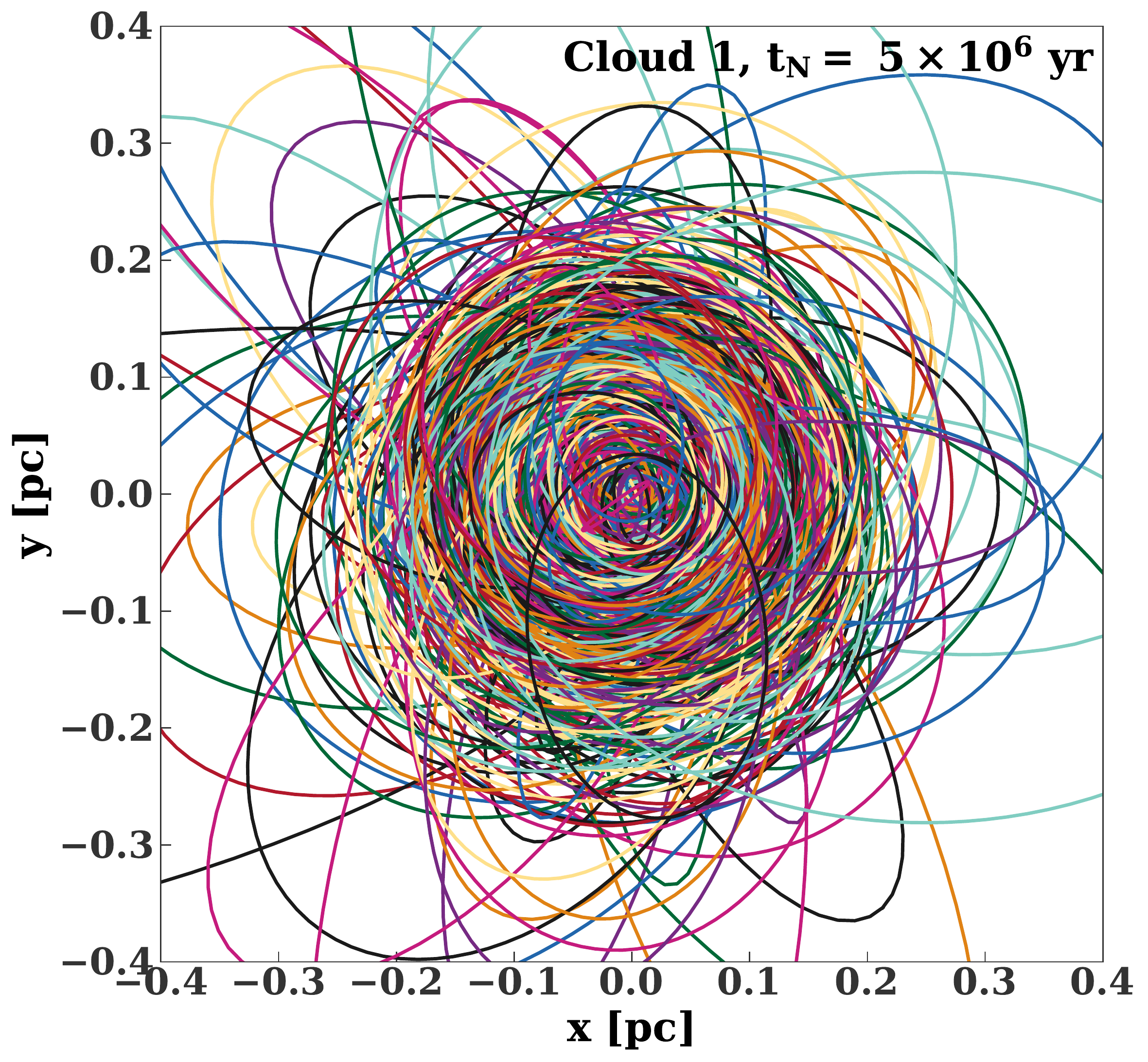}
    \includegraphics[width=\columnwidth]{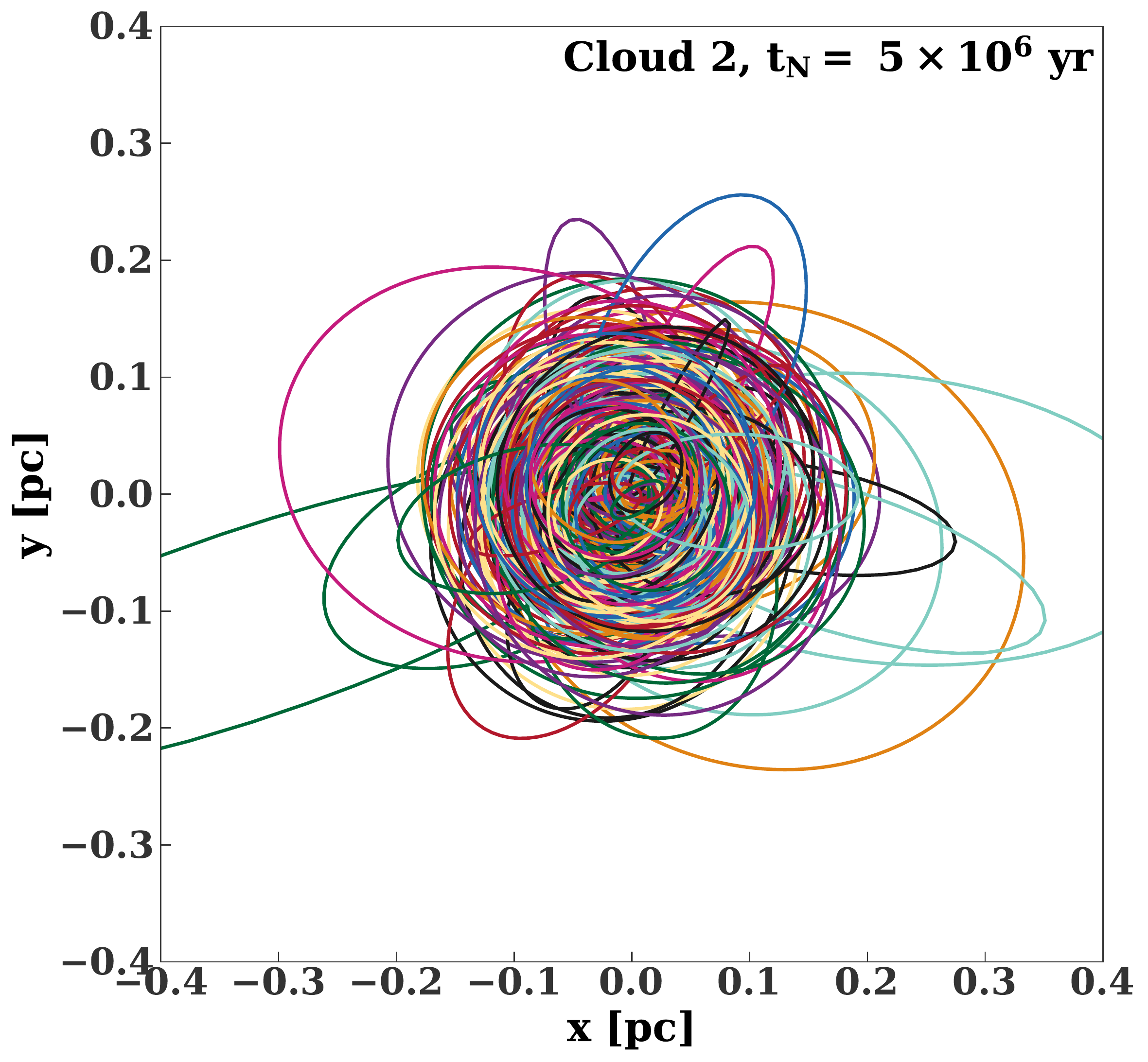}
    \caption{Initial (top) and final (bottom) orbits for the high resolution Cloud 1 (left) and Cloud 2 (right) simulations.}
    \label{fig:orbits}
\end{figure*}

Figure~\ref{fig:resComp}
shows the eccentricity and semimajor axis distribution (excluding disruptees) of the disks formed by Clouds 1 and 2 after 5 Myr of evolution in N-body simulations. After 5 Myr there is no longer any correlation between eccentricity and semimajor axis for Cloud 1, though a negative correlation is present for the Cloud 2 disc (i.e. orbits at smaller semimajor axes are on average more eccentric). 
Also, the disks have lost their apsidal alignment due to differential precession (see Figure~\ref{fig:orbits}).

As shown in Table~\ref{tab:ecc} and Figure~\ref{fig:resComp}, the semimajor axis and eccentricity distribution of the disk is dependent on resolution.
For example, the final mean eccentricity of the Cloud 1 stars is 0.48 in the low resolution simulation and 0.38 in the high resolution simulations. This is expected, as  the disc heats up via two-body relaxation, increasing its mean eccentricity \citep{richard_alexander+2007}. 

The relaxation time is proportional to 
$({N \langle m^2 \rangle})^{-1}$, where $N$ is the number of stars and $\langle m^2\rangle $ is the second moment of the mass function. We estimate the initial relaxation time at 0.1 pc for the discs formed from Cloud 1 and Cloud 2, in Table~\ref{tab:ecc}, using equation 3 from \citet{richard_alexander+2007}.\footnote{To accurately compute the relaxation time of a disk it is necessary to account for its flattened geometry, as done in this reference.} For comparison, we also show the relaxation time for a realistic $m^{-1.7}$ mass function, extending from $1$ to $100 M_{\odot}$ (for the disc masses in Table~\ref{tab:dis2}).\footnote{Interestingly, the combination $N \langle m^2\rangle$ is weak function of the minimum mass for an $m^{-1.7}$ mass function, changing by $\sim$15\% as the minimum of the mass function is decreased from $1 M_{\odot}$ to $0.1 M_{\odot}$ (at fixed disc mass).} The relaxation times for the high resolution simulations are within $40\%$ of the realistic estimates. Therefore, we use the high resolution simulations for observational comparisons. Also, we focus on low inclination stars for these comparisons, since high inclination stars would not be identified as members of the disc. Here, inclination is measured with respect to the mean of all the \emph{unit} angular momentum vectors. Low inclination orbits are those with inclinations less than the median ($\sim 13^{\circ}$ in high resolution simulations of both clouds after 5 Myr). Such orbits and the observed clockwise disc have a similar spread in inclination (e.g. \citealt{lu+2009} infer the half-opening angle of the disc to be $7\pm 2^\circ$.)

 The disc eccentricity inferred from observations is $0.27 \pm 0.07$ \citep{yelda+2014}, which is consistent with the mean eccentricity of low inclination stars in our high resolution simulations (see the third column of Table~\ref{tab:ecc}).

\begin{figure}
    \includegraphics[width=\columnwidth]{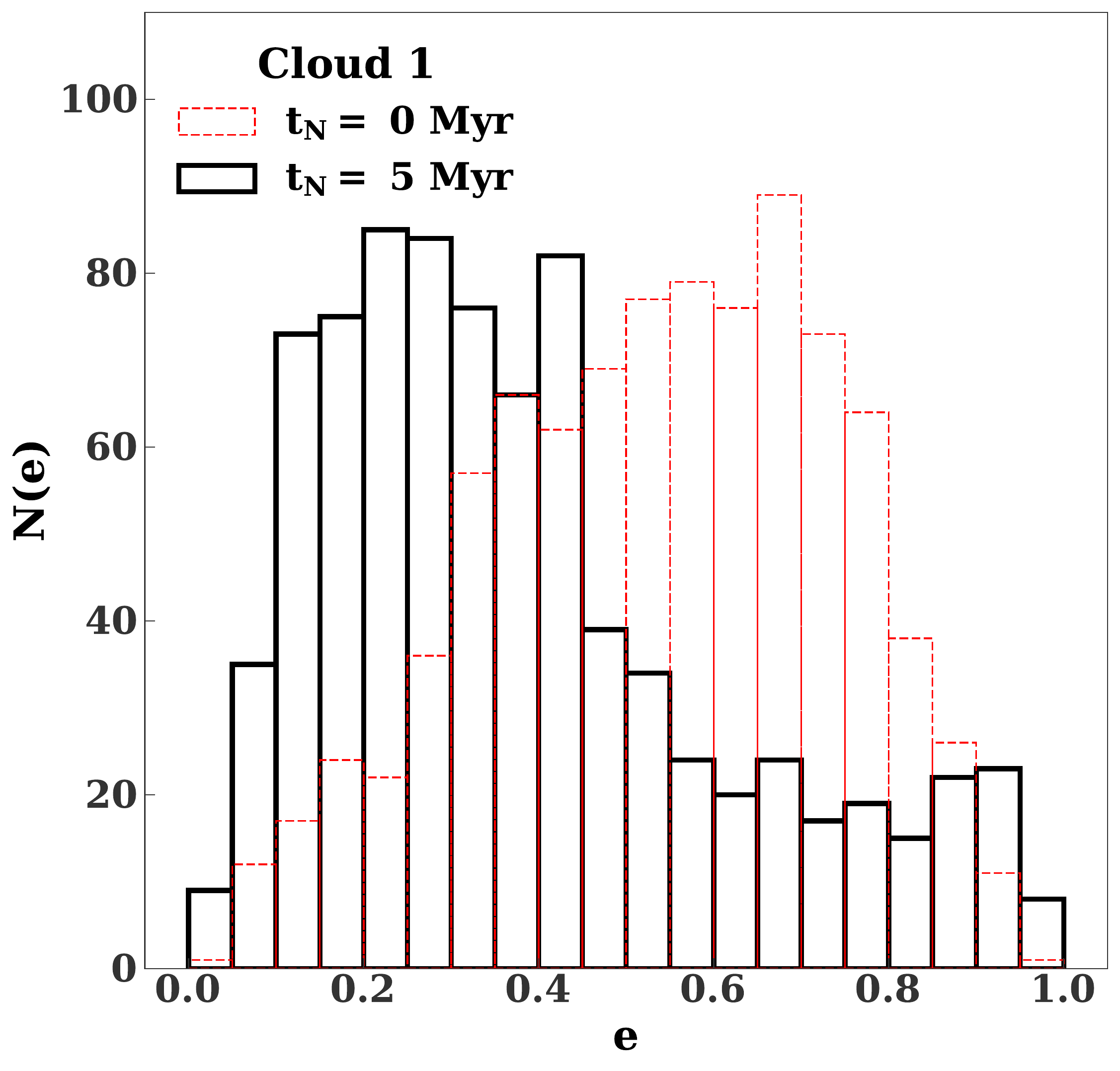}
    \includegraphics[width=\columnwidth]{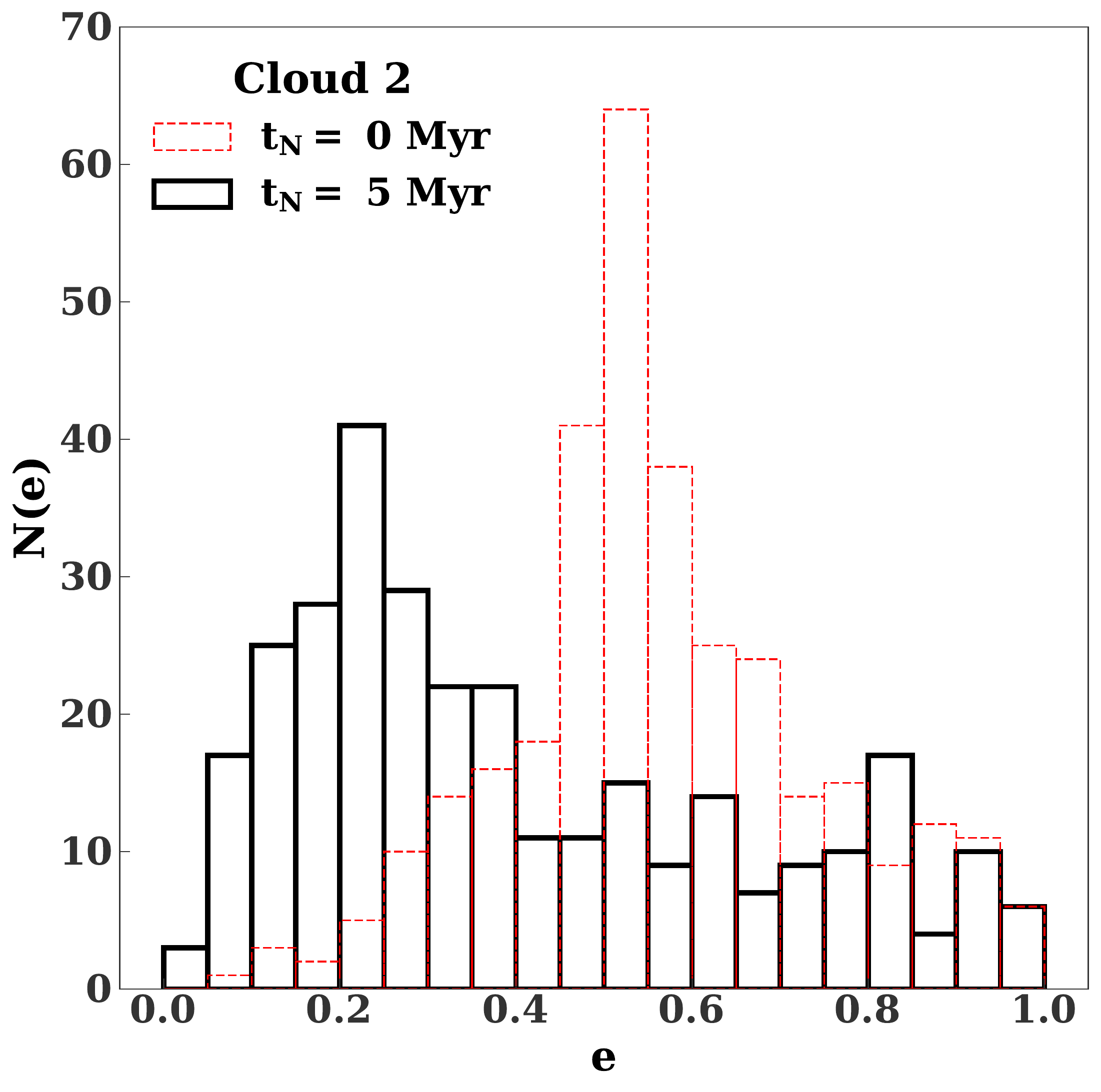}
    \caption{Initial and final eccentricity distribution in  (high resolution) $N$-body disc simulations initialized from Cloud 1 (top) and Cloud 2 (bottom). }
    \label{fig:e_before_after}
    \end{figure}

The eccentricity distribution of Clouds 1 and 2 disks after 5 Myr is unimodal like the eccentricity distribution inferred by \citet{yelda+2014} from observations (see Figure~\ref{fig:e_before_after}).
In contrast, in \citet{madigan+2009} and \citet{generozov&madigan2020}, the eccentric disk instability leads to a prominent bimodal structure. In fact, the shape of the final eccentricity distribution depends on the background density profile.  The final eccentricity distribution for the disc simulation with the flatter $r^{-1.1}$ background in \citet{generozov&madigan2020} also lacks a bimodal structure. In this paper we also assumed a relatively flat density profile (with a power-law index of $-1.16$), hence the bimodal eccentricity distribution is absent at late times. 

The bimodal structure is also time-dependent. At $\sim1$ Myr there is a second peak at high eccentricities. However, the eccentricity distribution only appears bimodal when high inclination stars are included. When these are filtered out, the peak at high eccentricities disappears. We find this is the case for various simulations, including those from \citet{generozov&madigan2020}.\footnote{In this case filtering out the high inclination stars results in a flat eccentricity distribution.}  This means we would not expect a bimodal eccentricity distribution to be observed within the disk, as high inclination stars would not be identified as disk members.

\begin{figure}
    \includegraphics[width=\columnwidth]{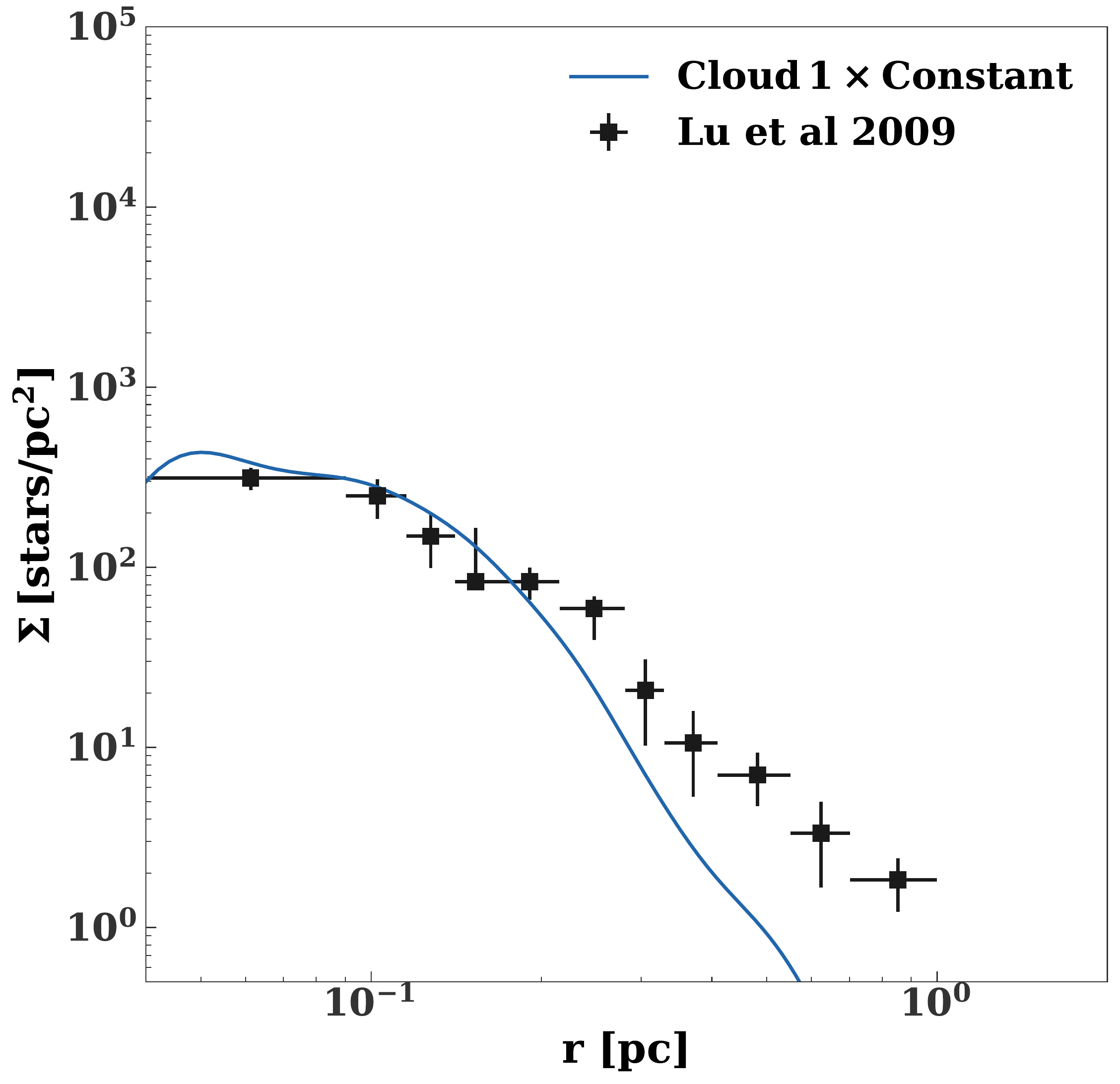}
    \includegraphics[width=\columnwidth]{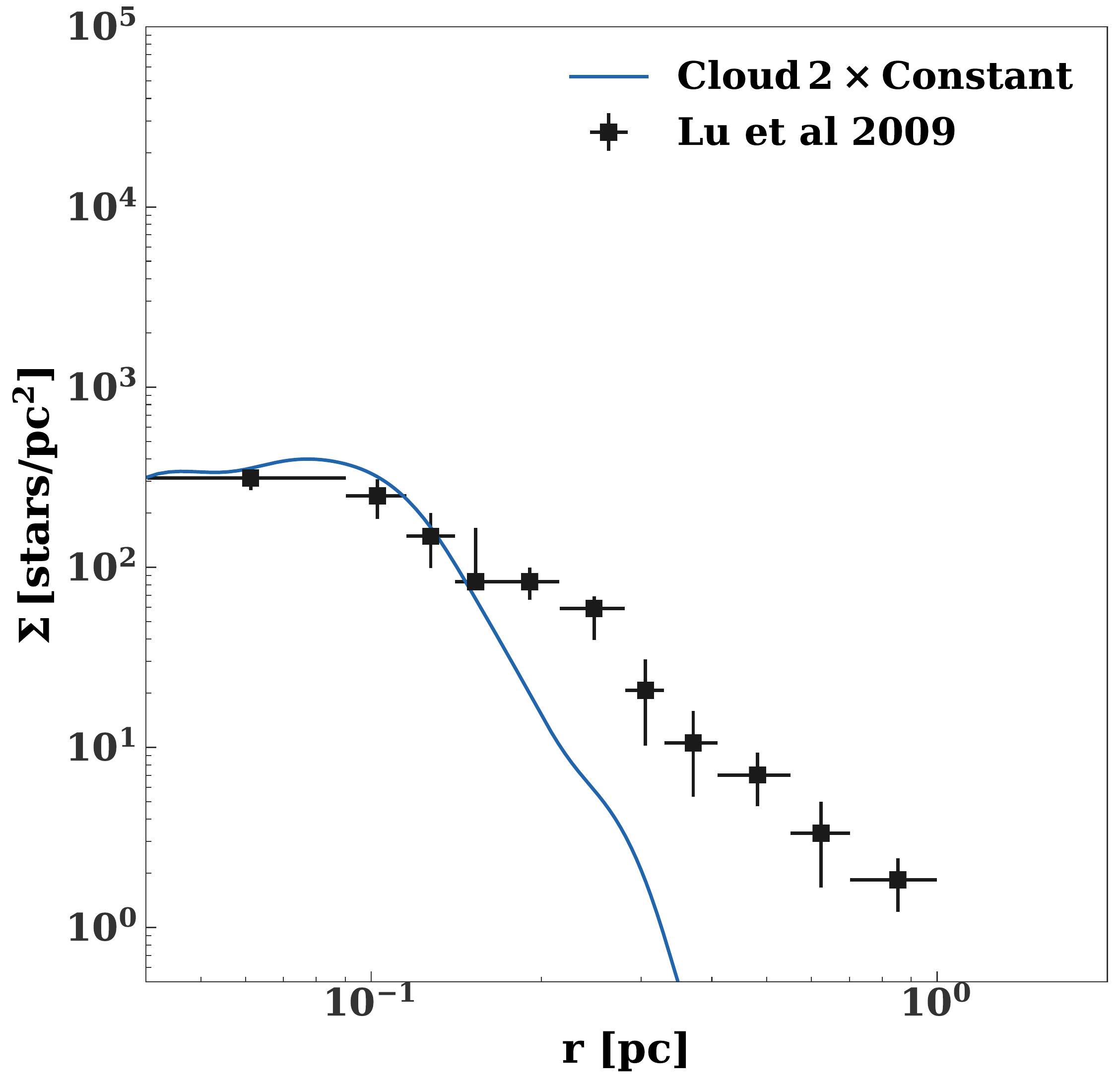}
    \caption{Surface density of low inclination stars after 5 Myr for Cloud 1 (\emph{top panel}) and Cloud 2 (\emph{bottom panel}), compared to the observed surface density profile from \citet{lu+2009}. The surface density from simulations is scaled by an arbitrary constant.}
    \label{fig:surf}
    \end{figure}

Figure~\ref{fig:surf} shows a comparison of the surface density profiles from Clouds 1 and 2 with the observed profile from \citet{lu+2009}. We scale the simulated profile by a constant to approximately match the overall normalization of the observed profile, considering the observations are incomplete at low stellar masses. The disks in our simulations are significantly more compact than the observed young disk(s) in the Galactic centre.

Overall, we are able to reproduce the mean eccentricity of the disk, but not its surface density profile. Also, in our simulation the inclination distribution of the disk becomes broader at small radii, whereas in the inclination distribution of observed clockwise disc becomes narrower at small radii \citep{naoz+2018}. However, our simulations lack several important physical effects as described in the following section and caution is warranted in interpreting such comparisons.

         \begin{table*}
                     \caption{\label{tab:ecc} Mean eccentricity after 5 Myr in low and high resolution Cloud 1 and 2 simulations.} 
            \begin{threeparttable}
            \begin{tabular}{cllll}
                Simulation  & <e>\tnote{1} & <e> (Low inc)\tnote{2}  & $t_{\rm rx}$\tnote{3} & $t_{\rm rx, real}$\tnote{4}  \\
                            &       &   & yr & yr \\
                \hline 
                \hline
                Cloud 1 (Low resolution)  & 0.48  & 0.46 & $3.0\times 10^4$ & $1.3\times 10^5$ \\
                Cloud 1 (High resolution) & 0.38 & 0.33 & $1.8\times 10^5$ & $1.3\times 10^5$\\
                Cloud 2 (Low resolution)  & 0.43  & 0.36 & $9.5\times 10^4$ & $1.8\times 10^5$ \\
                Cloud 2 (High resolution) & 0.40 & 0.31  & $2.5\times 10^5$ & $1.8\times 10^5$\\
                \hline
             \end{tabular}
             \begin{tablenotes}
             \item [1] Mean eccentricity of all bound particles in simulation
             \item [2] Mean eccentricity of all low inclination orbits--those within the median inclination. Inclination is defined with respect to the mean orientation of the disk.
             \item [3] Initial relaxation time of a disk annulus at 0.1 pc in the simulation, estimated from equation 3 in \citet{richard_alexander+2007}.
             \item [4] Estimate for initial disk relaxation time with a realistic mass function ($m^{-1.7}$, extending from $1$ to $100 M_{\odot}$).
             \end{tablenotes}
             \end{threeparttable}
         \end{table*}

\section{Discussion}

\label{sec:discuss}
\subsection{Additional effects}
In this section we discuss addition physical effects that could change the physical properties of the simulated disks, as well as observational effects that might bias the observed properties. 

    \begin{enumerate}
        \item Gas may persist for longer periods of time than we assume here, and circularize the stars via dynamical friction or gravitational torques. Alternatively, massive gas clumps could accelerate the two-body relaxation within the disk, which would increase the average eccentricity, and cause it to spread radially. 
        \item Star formation formation proceeds from the inside out in our hydrodynamic simulations (i.e. star formation first occurs close to the \sgra). Thus, some star formation may occur towards the outskirts of the stellar disk after the start of our N-body simulations. This could make the stellar disks less compact, and more consistent with observations.
        \item Our $N$-body simulations lack a mass spectrum. We would expect higher mass stars (that dominate the observed population) to have a lower average eccentricity and inclination than low mass stars due to two-body relaxation \citep{richard_alexander+2007}. Additionally, secular effects could lead to different mean eccentricities for different species \citep{foote2020,gruzinov+2020}, and lead to heavier species sinking towards the disk midplane \citep{szoelgyen&kocsis2018}. Finally, the heavier species would likely develop a steeper surface density profile than the light stars, potentially exacerbating the tension with the observed surface density profile.
        
        \item Unresolved binaries in the disc may bias the observed eccentricity distribution, as discussed in \citet{naoz+2018}.  
        
        We have used the procedure described in \citet{naoz+2018} to measure the apparent eccentricity distribution of Cloud 1 and Cloud 2 after 5 Myr, with a hypothetical, isotropic population of binaries. The orbital elements of the disc orbits are taken from our simulations, with the exception of the inclination and longitude of the ascending node, which are fixed to the observed mean values for the disc in the Galactic centre (130$^{\circ}$ and 96$^{\circ}$ respectively; \citealt{yelda+2014}). We assume the binaries consist of two $10 M_{\odot}$ stars on circular orbits of radius 0.05 or 0.5 au. 
        
        In the latter case the disc eccentricity distribution is virtually unaffected. In the former case, the mean disc eccentricity of bound stars for Cloud 1 increases from $\sim 0.38$ to $\sim 0.5$.
        Also, in this case $\sim 10\%$ of the stars appear to be unbound. Such a large population of unbound stars is not observed.
        
        \item We do not include the effects of discrete stars and compact objects in the nuclear star cluster surrounding the disk. As shown by \citet{perets+2018}, this can have a dramatic impact on its evolution, leading to flips in orientation, transient spiral arms, and warping.
        
        \item In \S~\ref{sec:bulkProp}, we compared the two-body relaxation time in our simulations to that expected with a realistic mass function. In that comparison we assumed a mass function extending to $100 M _{\odot}$. In reality stars above $\sim 40 M_{\odot}$ would have died over 5 Myr, increasing the two-body relaxation time by a factor of $\sim 3$. 
    \end{enumerate}

    Finally, \citet{madiganconf} previously found that the slope of the background stellar density can affect the number of disruptions coming from the disk. We explore this in the next section. 
    
\subsection{Slope of the stellar density profile}
    \label{sec:slope}
    We have assumed a relatively flat density profile with a power-law index of -1.16, based on \citet{schodel+2018}'s fits to the diffuse stellar light profile in the Galactic centre. In fact, the density profile may be significantly steeper especially on the scales of interest ($\sim 0.1$ pc).

    Fits to the resolved stellar population find a steeper profile close to $r^{-1.4}$ \citep{gallegocano+2020}. In fact, the diffuse light may be probing a somewhat different, younger stellar population that has not yet relaxed dynamically (see the discussion in Section 8.3 of \citealt{schodel+2020}).
    
    Theoretically, a relaxed stellar-density profile is expected to fall between $r^{-1.5}$ and $r^{-1.75}$ \citep{bahcall&wolf1976,alexander&hopman2009}. In fact, this relaxed cusp would only develop a factor of $\sim$5 times inside of the influence radius (or 0.4 pc in the Galactic centre -- see \citealt{vasiliev2017}). Thus, even if the power law index of the density profile is $-1.16$ near 1 pc, it would be significantly steeper on small scales for a relaxed population.
    Finally, the precession rate of the disc  depends on the total extended mass profile, including stellar mass black holes, which are expected to accumulate in the central region via mass segregation \citep{morris1993,miralda-escude&gould2000,freitag+2006, hopman&alexander2006a,merritt2010}. This population can form a very steep cusp. For example, previous work has found that the black holes can have  $r^{-2}$ over a large range of radii \citep{freitag+2006, alexander&hopman2009,aharon&perets2016,vasiliev2017}.
    We note that a significant population of stellar mass black holes is likely required to 
    reproduce the eccentricity distribution of the S-stars \citep{antonini&merritt2013,generozov&madigan2020}. Furthermore, observations of X-ray binaries in the Galactic centre also suggest the presence of a black hole cusp \citep{muno+2005,hailey+2018, generozov+2018}.

    To test the effects of a steeper density profile, we have added an additional external potential into (low resolution) simulations of Cloud 1 and Cloud 2, representing a black hole cusp. Specifically, we assume the black holes have an $r^{-2}$ density profile, with mass $M(r)=2.4\times 10^4 M_{\odot} \left(r/0.1 {\rm pc}\right)$ inside of radius $r$, as in \citet{antonini&merritt2013}.\footnote{This model can reproduce the S-star eccentricity distribution within these stars' lifetimes. However, assuming a single power-law index for the black hole profile is likely unrealistic: as black holes begin to dominate relaxation on small scales, their density profile should flatten \citep{vasiliev2017}.} The number of mean disruptions from Cloud 1 (Cloud 2) is 7.8 (5.6) with this steeper density profile. Thus, this black hole cusp does not strongly affect the results. We also tried a steeper, single component cusp with an $r^{-1.5}$ density profile and the same stellar mass inside of 1 pc ($\sim 1.2\times 10^6 M_{\odot}$). The mean number of disruption for Clouds 1 and 2 are 8.6 and 6.
    
    Previously, \citet{madiganconf} found that the slope of the background density profile strongly affects the disc  instability that pushes binaries to disruption. In order for disruptions to occur, the outer disc  has to precess ahead of the inner disk. As the stellar density becomes flatter, differential precession and disruptions are suppressed.
    
\begin{figure}
        \includegraphics[width=0.99\columnwidth]{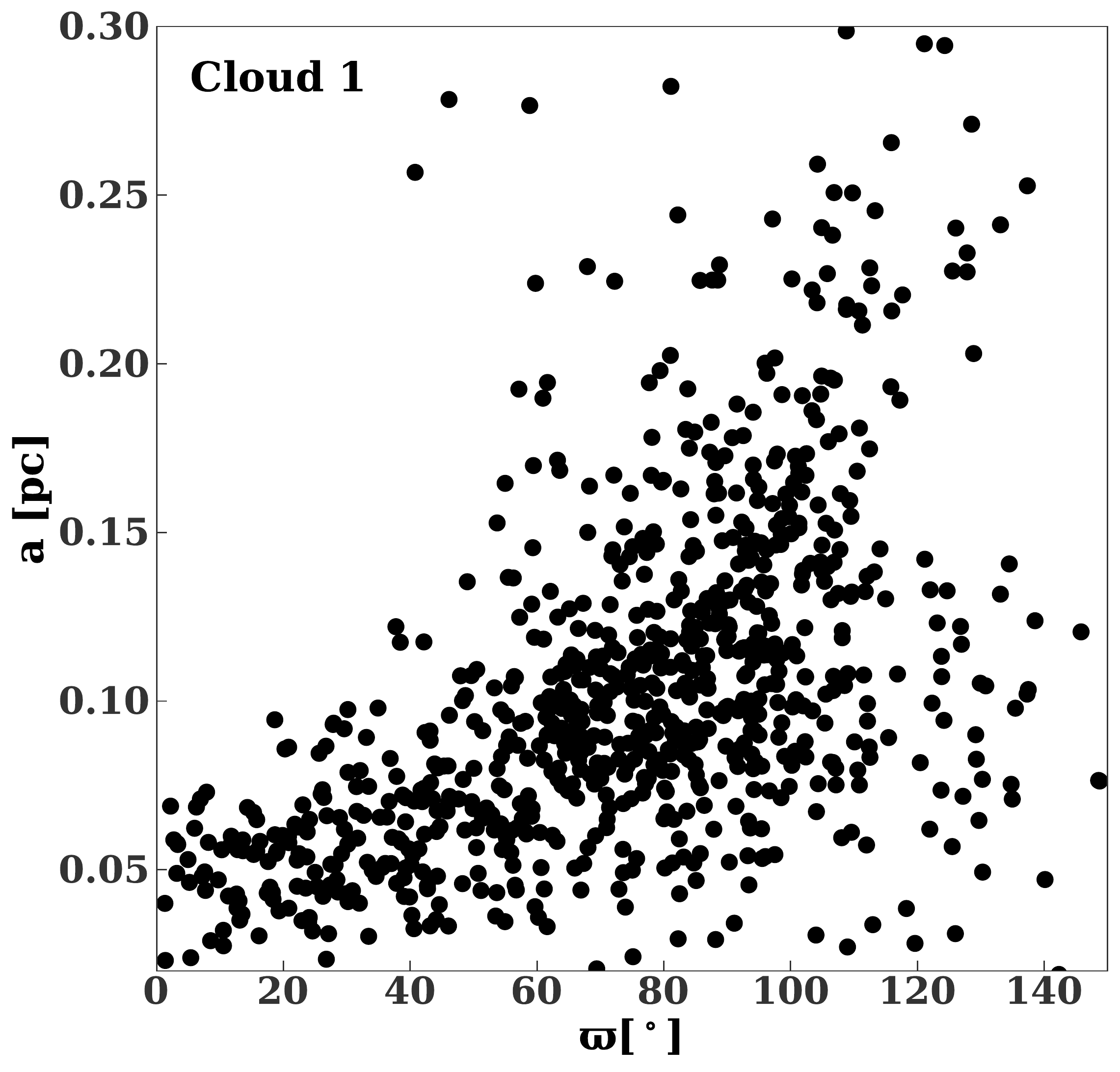}
        \includegraphics[width=0.99\columnwidth]{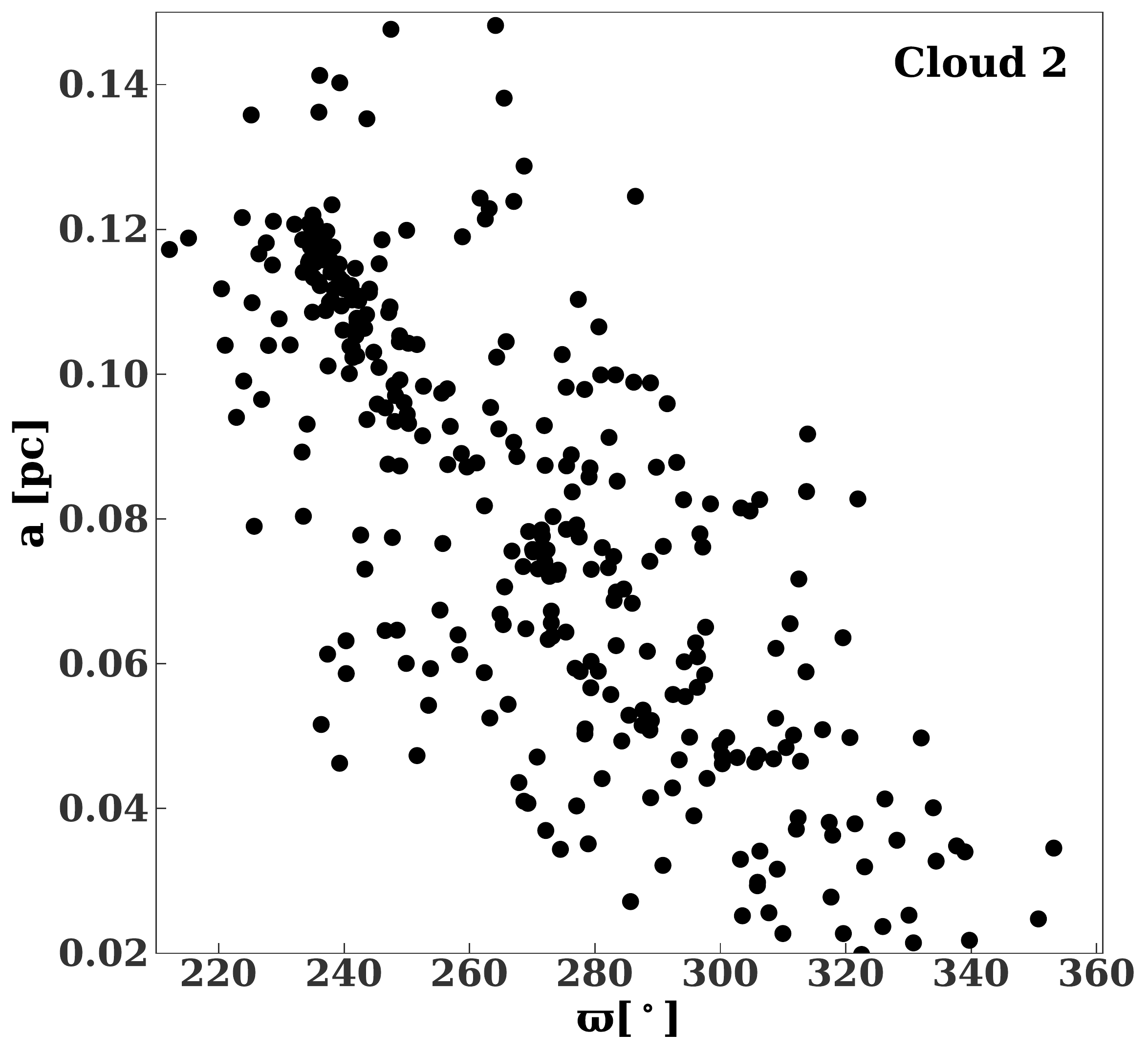}
        \caption{Initial orientation of stellar orbits in Cloud 1 (\emph{top panel}) and Cloud 2 (\emph{bottom panel}) as a function of semimajor axis. Cloud 1 (2) has its angular momentum along the negative (positive) z-axis, and precesses towards increasing (decreasing) $\varpi$. Thus, the disk will precess towards the right (left) in the top (bottom) panel.} 
        \label{fig:twist}
    \end{figure}

    However, in our case the disks start with a strong primordial twist as illustrated in Figure~\ref{fig:twist}. For both Cloud 1 and Cloud 2 the bulk of the outer disc starts significantly ahead in its precession. This twist allows the eccentric disc instability to develop even in the presence of a steep cusp. 
    
    We recover a bimodal eccentricity distribution for Cloud 2 with the $r^{-1.5}$ background density profile. However, the bimodal structure is still absent for Cloud 1. This suggests that a steep background density profile is a necessary, but insufficient condition, for producing a bimodal eccentricity at late times. Cloud 1 produces the most massive stellar disk, with the fastest two-body relaxation. This enhanced two-body relaxation may be suppressing the peak at high eccentricities, by causing stellar orbits to thermalize.
    
    The numerical experiments described in this subsection are not self-consistent. Although we changed the background stellar density in our $N$-body simulations, we are still using snapshots from hydrodynamic simulations with a flatter background density profile as initial conditions. Thus, the twist in Figure~\ref{fig:twist} may be an artifact of strong differential precession induced by this flatter density profile. However, the initial conditions here are somewhat arbitrary to begin with, as discussed in \S~\ref{sec:ic}. It is possible that a strong primordial twist could also be seeded by a non-uniform cloud, even with a steep background density profile.

 \subsection{Single star disruptions}
\label{Nstar}

    We can infer the number of single star disruptions from our disc simulations. For a $10 M_{\odot}$ star the 
    tidal disruption radius is $\left(M_{\rm bh}/M_*\right)^{1/3} R_*\approx 6.3\times 10^{-6}$ pc. We find that $\sim 2-3\%$ of 
    particles reach a smaller pericentre for both clouds. If the particles are in fact binaries, they would be tidally disrupted before their component stars could reach their tidal disruption radii, since they would be in the empty loss cone regime \citep{generozov&madigan2020}. Thus, the
    single star disruption rate is $\sim 2\% (1-f_{\rm bin})$, where $f_{\rm bin}$ is the binary fraction in the disc.

  \subsection{Properties of ejected stars}
    Recently \citet{koposov+2019} discovered a hypervelocity star that was ejected from the Galactic centre 5 Myr ago. Such stars would be produced by our disks, since binary disruptions typically eject one star from each disrupted binary from the Galactic centre \citep{hills1988}.

\begin{figure}
    \includegraphics[width=\columnwidth]{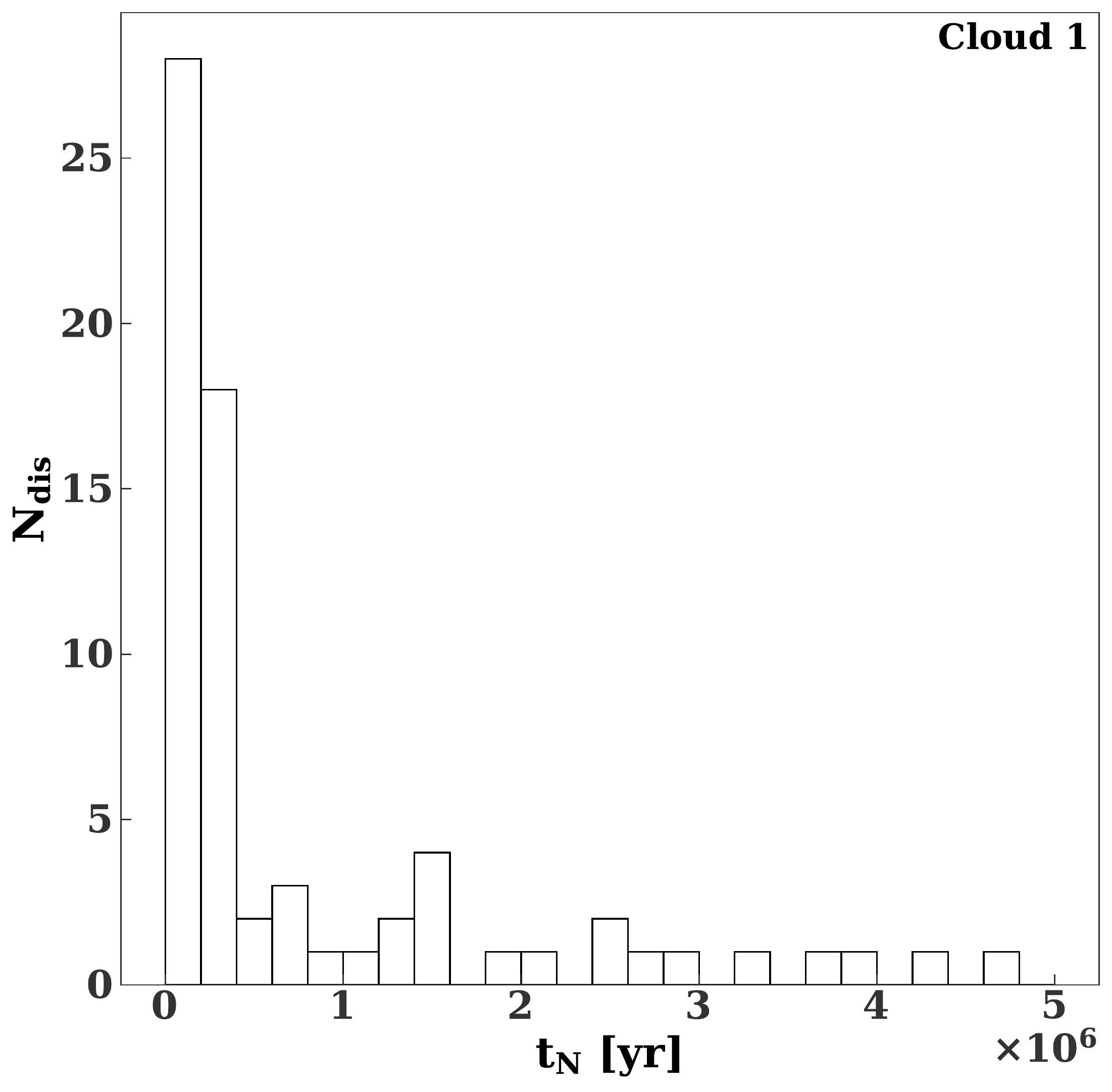}
    \includegraphics[width=\columnwidth]{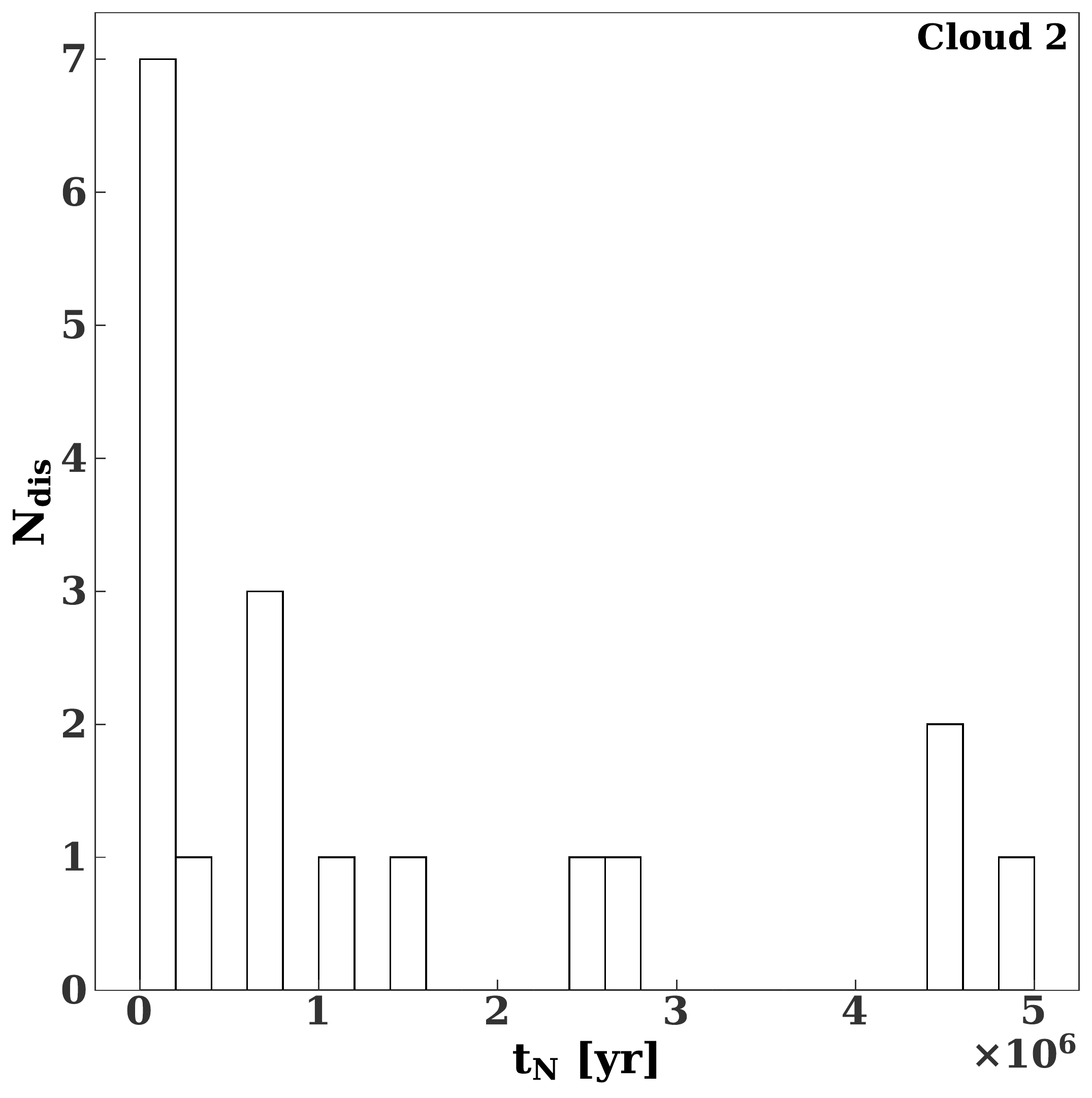}
    \caption{Distribution of disruption times for $N$-body disc simulations initialized from Cloud 1 (top) and Cloud 2 (bottom). These results are from our high resolution simulations.}
    \label{fig:dis_times}
\end{figure}

\begin{figure*}
    \includegraphics[width=\textwidth]{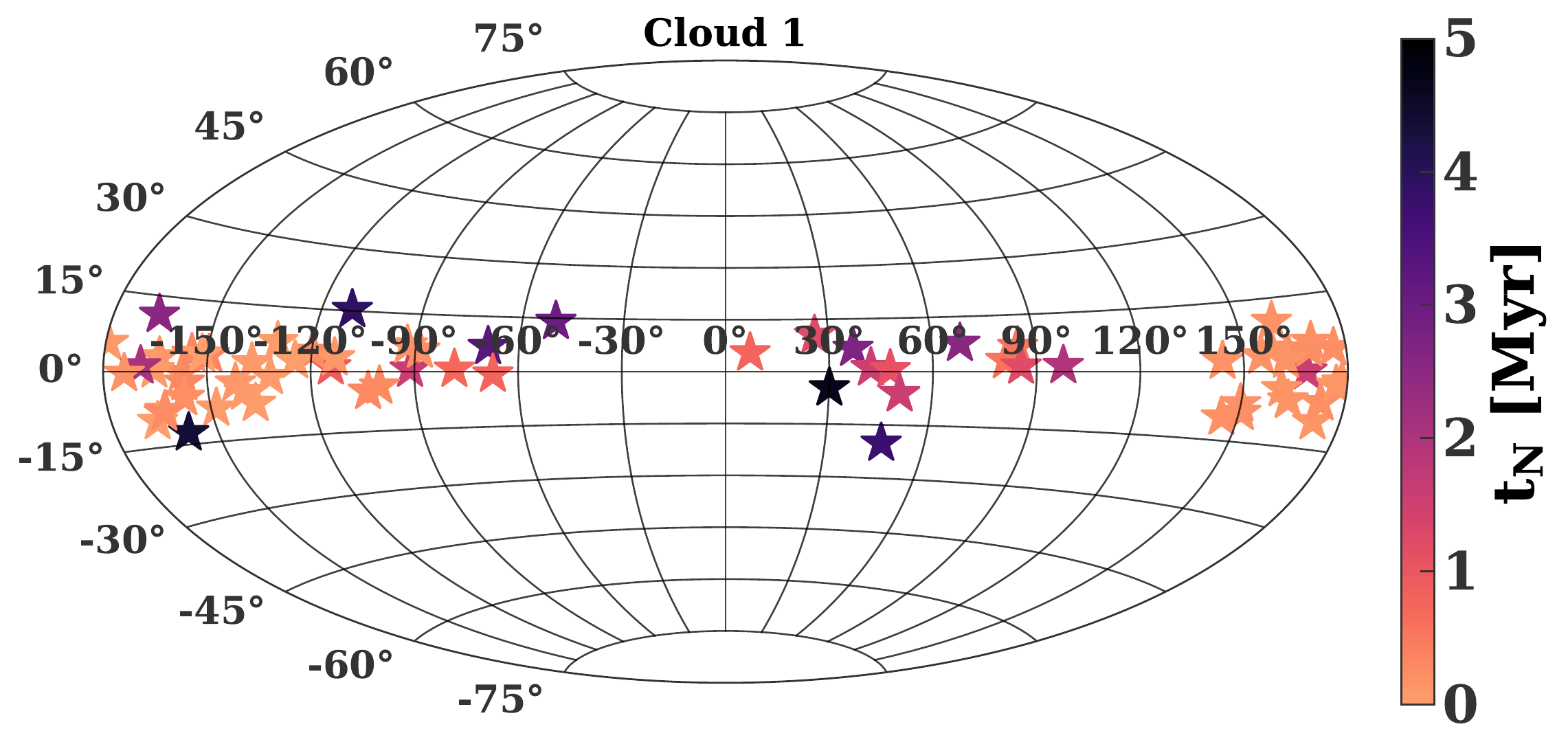}
    \includegraphics[width=\textwidth]{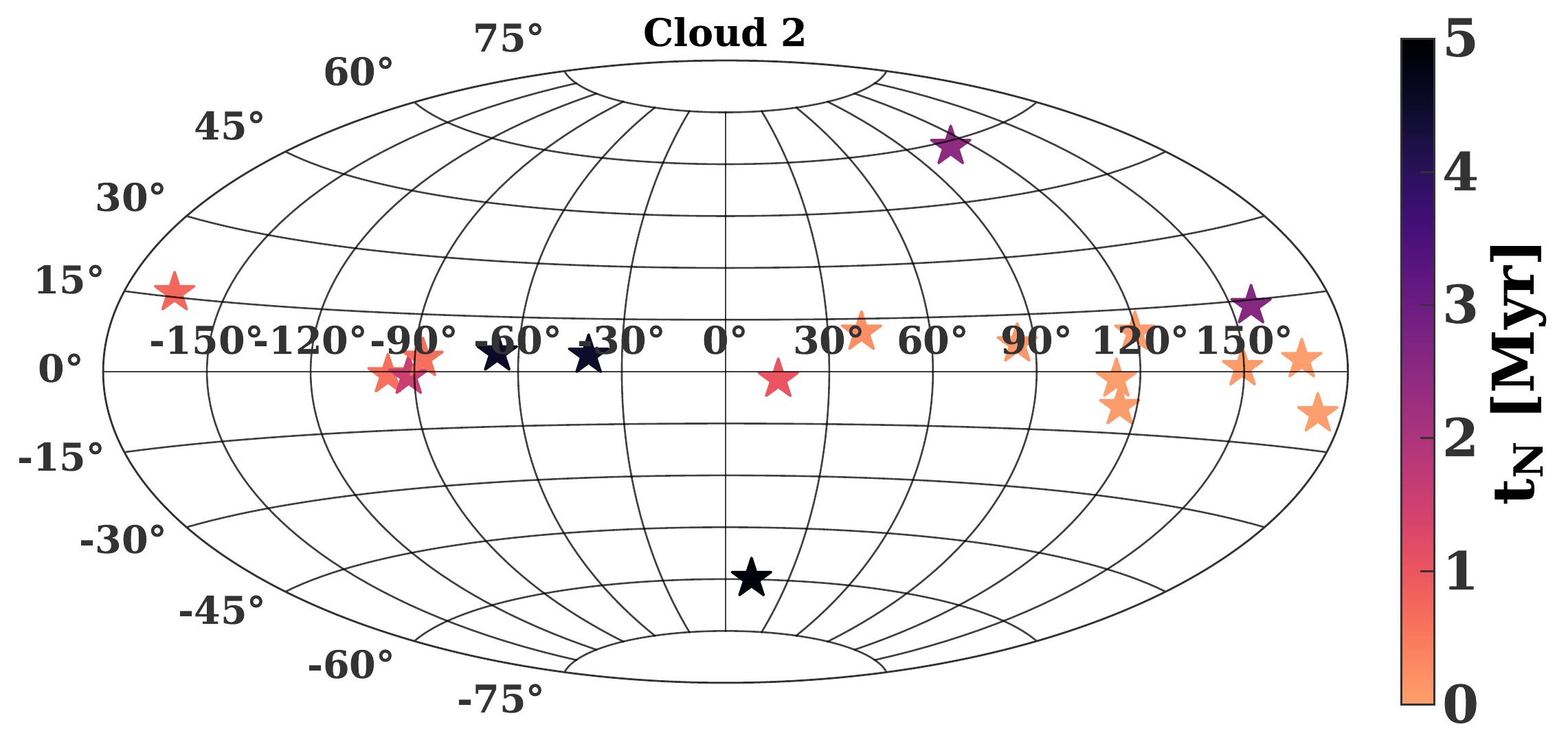}
    \caption{Aitoff projection of polar angles of `ejected stars'' velocity vectors. Here the equator corresponds to the disk midplane. The top (bottom) panel corresponds to $N$-body disc simulations initialized from Cloud 1 (Cloud 2). Each point is colored by the time disruption (and ejection) would occur.}
    \label{fig:dis_angs}
\end{figure*}

     As shown in Figure~\ref{fig:dis_times}, most disruptions occur within the first Myr for both clouds, though there is a tail extending to late times. 
    We follow the procedure described in \citet{generozov2020} to predict the angular distribution of hypervelocity stars produced by each cloud. In particular, for each disruptee, we initialize a hyperbolic orbit with the same orbital angles and pericentre, and assume a fixed semimajor axis of -0.01 pc.\footnote{The orbital geometry depends on the ratio of pericentre to semimajor axis, which is approximately $\left(M_{\rm bh}/m_{\rm bin} \right)^{-1/3}$. This would have a broader range than assumed here. However, a realistic spread in this ratio does not significantly affect the angular distribution in Figure~\ref{fig:dis_angs}.} We then measure the velocity at infinity for this orbit. Figure~\ref{fig:dis_angs} shows an aitoff projection of the polar angles for these velocities. Stars ejected by Cloud 1 show weak clustering in azimuth. This will lead to a concentration of young hypervelocity stars in a particular region of the sky. For stars ejected within first 1 (5) Myr the p-value that the azimuthal angle is drawn from a uniform distribution is $3\times 10^{-4}$ (0.002) according to the Kolmogorov--Smirnov test. The circular standard deviation of the azimuthal angle is $47$ ($69$) degrees. For comparison, in the simulations of \citet{generozov&madigan2020} and \citet{generozov2020}, the standard deviation of stars ejected in first Myr was $30-60$ degrees.
    
    The production rate of hypervelocity stars may be dominated by other channels like scattering by molecular clouds from large scales \citep{perets+2007, perets&gualandris2010}. Alternatively, some background stars in the Milk Way's nuclear star cluster may be excited to disruption by the disc. In fact, S5-HVS1 could be such a star, considering its spectroscopic age (25-93 Myr; \citealt{koposov+2019}) exceeds the disc's
    
\subsection{Differences with previous work}
\label{sec:litComp}

\citealt{gualandris+2012} performed high accuracy N-body simulations initialised from hydrodynamical simulations of molecular cloud infall and star formation by \citealt{Mapelli2012}. In many regards our work is similar to that of the quoted authors, but results do bear quantitative differences. In particular, \citealt{Mapelli2012} found discs that are significantly less eccentric ($e\sim 0.2-0.4$). In contrast, the mean eccentricity of the discs at the start of our N-body simulations is $\sim 0.5$ (with a maximum eccentricity up to 0.99).

The disparity in the initial eccentricity distribution may be driven by differences in the initial conditions, physical problem setup, and numerical techniques. In particular, the biggest differences are:
\begin{enumerate}
    \item The \citealt{Mapelli2012} gas cloud is 1.9 times larger; it starts at a galactocentric distances of 25 pc, whereas we focus on clouds that start at 10 pc. The initial gas velocity is the same everywhere in our clouds while the former authors seed their clouds with supersonic turbulence.
    \item The background stellar density profile is different. \citet{Mapelli2012} assume an $r^{-2}$ ($r^{-1.4}$) density profile outside (inside) of 0.39 pc, whereas, we use the broken power-law in equation~\eqref{eq:schodelFit}.
    \item Our clouds are on less radial orbits. The initial pericentre of the centre of our clouds is $\sim0.2$ pc. In  \citet{Mapelli2012} the impact parameter of the cloud centre is 0.01 pc (and the pericentre would be even smaller). 
    \item The thermodynamics of gas in the simulations is different. We impose a minimum (irradiation) temperature of 15 K, whereas \citealt{Mapelli2012} impose a much higher minimum temperature of 100 K. These temperatures are all well below the local virial temperature, $\sim 10^6 -10^7$~K, so these choices are probably immaterial for the large scale cloud infall dynamics but may affect the dynamics of gas and the stellar mass function in the disc. Radiative cooling is included in both our hydrodynamic simulations and the one used by \citet{gualandris+2012}.
    \item The artificial viscosity of the SPH simulations in the two studies is different. The latter detail may affect shocks that form during the cloud infall and subsequent disc formation.
    \item Another significant difference is the prescription for star formation. While we introduce sink particles that are allowed to grow by further gas accretion, \citealt{Mapelli2012} do not introduce sink particles. 
\end{enumerate}
Currently, it is not possible to pinpoint which of these factors drives the differences in our results. 

Overall, more simulation work is needed to ascertain the robustness of our \citep[and][]{gualandris+2012} conclusions, with attention to both numerics and the uncertainties in the stellar potential and the initial configuration of the molecular cloud prior to its infall in the Galactic Centre. In particular, the effect of the initial cloud orbit warrants attention, as counter-intuitively we obtained a more eccentric disc with a less radial orbit.

\section{Summary}
\label{sec:sum}
In this paper, we have simulated disruptions of molecular clouds in the Galactic centre. Such disruptions can lead to the formation of the lopsided, eccentric discs that would explain the young disk observed within the central parsec of the Galaxy. Moreover, these disks are subject to a secular gravitational instability that can excite binary stars in the disk to extreme eccentricities and disruption. The remnants from these disruptions would explain the S-star cluster within $\sim$0.04 pc of the Galactic centre.
Our main conclusions are summarized as follows 

\begin{enumerate}
    \item The velocity of the initial cloud can be constrained from observations. Clouds with velocities $\gsim$100 km s$^{-1}$ at 10 pc form long star-forming filaments that are not observed in the Galactic centre. Therefore such high velocities can be ruled out. Additionally, velocities $\lsim 20$ km s$^{-1}$ can be ruled out, as they produce disks that are too small in radial extent to explain the observed clockwise disk. The most promising cloud velocities are $\sim 40-80$ km s$^{-1}$.
    
  Similarly the mass of the cloud can be constrained. Clouds with masses between a few$\times 10^4 M_{\odot}$ and $10^5 M_{\odot}$ are the most promising. Lower mass clouds produce too few stars to explain the observations. Higher mass clouds would likely produce too many.
    
    \item We identified two promising Cloud initial conditions for reproducing the young clockwise disk in the Galactic centre, as well as the S-stars. Overall, these clouds can push $\sim 6-8\%$ of their constituent binaries to disruption. 
    
    \item We are able to reproduce the mean eccentricity of the clockwise disk. The final eccentricity profiles in our simulations is unimodal, although a bimodal distribution is present at early times. Previous work (\citealt{madigan+2009,generozov&madigan2020}) found a bimodal eccentricity distribution due to a steeper background density profile.

    \item The eccentricity distribution is inclination dependent. For example, even if a bimodal eccentricity distribution is present, it will only show up if high inclination stars are included in the distribution. As such stars are not identified as disk stars in observations, we would not expect a bimodal eccentricity in the observed clockwise disk. 
    
    \item We are not able to reproduce the observed surface density profile of the clockwise disk. In particular, the surface density falls off too steeply with radius. 
    
    \item There are a number of simplifications we have made that may affect the aforementioned observational comparisons. In particular, our N-body simulations lack a mass spectrum, a live nuclear star cluster, and gas.
    
\end{enumerate}

We have demonstrated that the tidal disruption of a molecular cloud is a viable way to explain the young disc stars in the Galactic centre, as well as the S-stars and hypervelocity stars.

The eccentricity distribution of non-disc S-stars can be used as a test of our model. In particular, we predict young stars outside of the disk should have higher average eccentricities.  Additionally, the presence or absence of counter-rotating discs as claimed in \citet{genzel+03, paumard+2006} (though see \citealt{lu+2009,yelda+2014}) can be used as an additional constraint on the initial conditions.

\section*{Acknowledgements}
We thank the anonymous referee for helpful comments.

We thank Andreas Eckart, Stefan Gillessen, Yuri Levin, Avi Loeb, Hagai Perets, and   Rainer Sch{\"o}del for helpful comments.

AM gratefully acknowledges support from the David and Lucile Packard Foundation. 
This work utilized resources from the University of Colorado Boulder Research Computing Group, which is supported by the National Science Foundation (awards ACI-1532235 and ACI-1532236), the University of Colorado Boulder, and Colorado State University. S. N. acknowledge the funding from the UK Science and Technologies Facilities Council, grant No. ST/S000453/1. This work made use of the DiRAC Data Intensive service at Leicester, operated by the University of Leicester IT Services, which forms part of the STFC DiRAC HPC Facility (www.dirac.ac.uk).

\section*{Data Availability}

Data used in this study will be made available upon reasonable request.



\bibliographystyle{mnras}
\bibliography{master.bib} 





\bsp	
\label{lastpage}
\end{document}